\documentclass[%
 reprint,
 amsmath,amssymb,
 aps,
 pra,
]{revtex4-2}

\usepackage{graphicx}
\usepackage{dcolumn}
\usepackage{bm}
\usepackage{bbold}
\usepackage{amssymb}
\def\Xint#1{\mathchoice
   {\XXint\displaystyle\textstyle{#1}}%
   {\XXint\textstyle\scriptstyle{#1}}%
   {\XXint\scriptstyle\scriptscriptstyle{#1}}%
   {\XXint\scriptscriptstyle\scriptscriptstyle{#1}}%
   \!\int}
\def\XXint#1#2#3{{\setbox0=\hbox{$#1{#2#3}{\int}$}
     \vcenter{\hbox{$#2#3$}}\kern-.5\wd0}}

\def\dashint{\Xint-}

\usepackage{hyperref}
\hypersetup{
	colorlinks,%
	citecolor=blue,%
	filecolor=blue,%
	linkcolor=blue,%
	urlcolor=blue
}

\begin{document}

\preprint{APS/123-QED}

\title{Temporally correlated quantum noise in driven quantum systems}

\author{Balázs Gulácsi}
 \email{balazs.gulacsi@uni-konstanz.de}
\author{Guido Burkard}%
 \email{guido.burkard@uni-konstanz.de}
\affiliation{%
 Department of Physics, University of Konstanz, 78457 Konstanz, Germany
}

\date{\today}
         

\begin{abstract}
The ubiquitous effects of the environment on quantum-mechanical systems generally cause temporally correlated fluctuations. This particularly holds for systems of interest for quantum computation where such effects lead to correlated errors. 
The Markovian approximation neglects these correlations and thus fails to accurately describe open-system dynamics where these correlations become relevant.
In driven open systems, yet another approximation is persistently used, often unknowingly, in which one describes the decay effects independently from the time-dependent controlling fields acting on the system, thereby ignoring further temporally correlated effects.
To overcome these shortcomings, we develop a quantum master equation for driven systems weakly coupled to quantum environments that avoids the aforementioned field-independent approximation, as well as the Markovian approximation. Our method makes it possible to track all occurring decay channels and their time-dependent generalized rates which we illustrate in the example of a generally driven two-level system. We also demonstrate that correlated and field-dependent dissipative effects can lead to an increase in the performance of single-qubit gate operations.
\end{abstract}

\maketitle


\section{\label{sec:level1}Introduction}

Harnessing quantum phenomena holds immense potential to revolutionize real-world technologies, including quantum sensing \cite{sensing}, quantum information processing \cite{nielsen,qip,Wendin2017,scqubit}, cryptography \cite{crypto1,crypto2}, and quantum communication \cite{qcomm}. However, the practical implementation of these ideas faces a formidable challenge in the form of decoherence. Realistic quantum systems are never completely isolated from their environment, and the system-environment interactions tend to destroy the coherence between the states of the reduced quantum system \cite{decoh}. 
An essential part of these technologies is the precise control of a quantum system. For instance, quantum computation can be viewed as the art of controlling the time evolution of highly complex, entangled quantum states in physical hardware registers \cite{Wendin2017}. This control is associated with the time dependence in the Hamiltonian describing the actual quantum processor. As such, a thorough understanding of decoherence mechanisms acting on driven quantum systems is imperative for the advancement of quantum technologies into everyday applications.


The influence of the environment is usually captured using the theory of open quantum systems \cite{breuer}. Within this framework, the environment as a noise source is described as a large quantum system that interacts with the system of interest. The dynamics of the relevant system are then obtained by tracing out the environmental degrees of freedom, which leads to a quantum master equation. In general, such a description always leads to non-Markovian evolution \cite{nonloc,Rivas2014}. To render the problems tractable, simplifications are made, such as the Born-Markov and secular approximations \cite{GORINI1978,semigroup} leading to the well-known Lindblad equation \cite{lindblad,gks}. Broadly speaking, the Lindblad equation extends the von Neumann equation, which describes the dynamics of the closed system, with dissipators that model the effects of the environment. While this is effective in capturing the essential dynamics of open quantum systems, it overlooks crucial aspects of the system-environment interaction, particularly the role of correlations, leading to inaccuracies in predicting the behavior of quantum systems in realistic scenarios \cite{corr1,burst,corr2,ulm1,Lidar1,Agarwal2024,PRXiqm}.
\begin{figure}[b]
    \centering
    \includegraphics[width=0.6\linewidth]{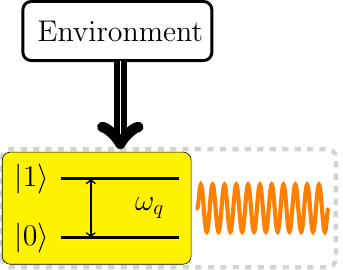}
    \caption{Decoherence acting on a driven system. The effect of the environment on the relevant quantum system (yellow box) is dependent on the time-dependent manipulations (orange wavy line) by which the quantum hardware is controlled. 
Within the \emph{field-independent} approximation, the line between control and environment is blurred and one can imagine the decoherence acting independently from the controlled self-evolution of the system.}
    \label{fig:uncorr}
\end{figure}

For driven quantum systems, another predominant approximation is made for the description of decoherence. Within this approximation, the presence of a time-dependent control field affects the dynamics only in the Hamiltonian contribution of the master equation, and the additional dissipative contributions remain field-independent  (see Fig.~\ref{fig:uncorr}).
For instance, a phenomenological open-system model of qubits in a quantum register is described using two main processes. A qubit may relax to its ground state with rate $1/T_1$, with $T_1$ being its relaxation time. It may also undergo dephasing with a rate $1/T_\varphi$. Using standard measurements for the relaxation times $T_1$ and $T_2$, the rates can be determined using the formula $1/T_2=1/2T_1+1/T_\varphi$. When the qubits are controlled, the von Neumann equation with the full time-dependent Hamiltonian of the quantum processor is conveniently extended with the dissipators describing the relaxation and dephasing processes using the relaxation times as parameters. 
The flaw of this approximation arises from the measurements employed to determine $T_1$ and $T_2$, which stem from experiments conducted on the \emph{undriven} system and assuming decoherence acts the same way in the presence of a drive. This is a drastic simplification of the effects of quantum noise acting on driven systems. On the contrary, to reduce deleterious environmental influence, a more accurate understanding is required. Moreover, in the context of quantum computation,  computational timescales are very much restricted by decoherence, and thus there is a need for rapid operations on quantum systems which translates to stronger driving fields. Within these conditions, the dissipative effects cannot be treated independently from the time-dependent manipulations that are applied to the system \cite{strong,bridging}. 
Nevertheless, this field-independent approximation finds widespread application and it is often the go-to description of decoherence on driven quantum systems \cite{Ex0,Ex1,Ex2,Ex3,Ex4,Ex5,Ex6,Ex7,Ex8,noisygates2,fastgate}. 


The interest in the environmental effects on driven quantum systems dates back to the late 1950s. In the context of nuclear magnetic resonance and relaxation, the main focus of study was on nuclear spins interacting with each other, as well as the molecules of the surroundings and with the externally applied time-dependent magnetic fields. The developed methods avoid the field-independent approximation; however, they are restricted to cases when the time-dependence of the driving field can be eliminated by a unitary transformation with the introduction of a suitable rotating frame \cite{bloch,oldies1,oldies2,oldies3}. Furthermore, the special case of periodic driving fields enables the use of the Floquet formalism \cite{GRIFONI1998}. Other, more modern approaches also rely on certain limitations of the properties of the environment or the applied field. In particular, the interplay of swiftly decaying correlations in the environment and the slowly changing external fields leads to adiabatic quantum master equations \cite{Amin2009,adiabaticme,Vega2017}. Within this framework, decoherence acts on the instantaneous eigenstates of the time-dependent Hamiltonian. Later works extended this theory to non-adiabatic regimes by realizing that the Markovian assumption entails much more than the adiabatic assumption on the time dependence of the reduced system \cite{BM1,adiabatic1}. 


In this paper, we aim at the construction of a general master equation that does not rely on any approximation other than the weak coupling assumption between system and environment. In Sec.~\ref{sec:level2}, we first discuss the Hamiltonian of a general driven system coupled to multiple environments, and we introduce a suitable picture in which we utilize the time-convolutionless projector method to obtain the master equation which accounts for the time-dependence of the closed system. Working in the standard basis of the reduced system, we can derive all the conventional dissipative channels valid for the undriven system, along with the additionally arising decay channels due to the drive and their time-dependent rates. We consider these as the main results of this paper. In Sec.~\ref{sec:level3}, we illustrate our method and its consequences on a generally driven two-level system. In Sec.~\ref{sec:level4}, we apply our model to analyze the realizations of single-qubit gates implemented for weakly anharmonic Josephson qubits. Finally, in Sec.~\ref{sec:level5} we conclude our work.

\section{\label{sec:level2}Theory of weak decoherence in driven quantum systems}

We wish to model the perturbative effects of the environment acting on a general time-dependent quantum system representing, e.g., a piece of a quantum processor,
\begin{eqnarray}
H=H_0+V(t)+H_B+H_{SB},
\label{eq:ham}
\end{eqnarray}
where $H_0$ represents the quantum system whose Hilbert space contains the computational subspace in the case of a quantum processor. The operator $V(t)$ incorporates the control of the system, typically this is described semiclassically by time-dependent external fields that couple to system operators representing system degrees of freedom, such as charge or spin. The environment is present through $H_B$, embodying a large collection of uncontrollable degrees of freedom that interact with our system of interest through the term $H_{SB}$. In the following, we make the weak coupling assumption so that the system-environment interaction is linear and weak compared to all other energy scales in the problem. These entail the form of the interaction term to be
\begin{eqnarray}
\label{eq:int}
H_{SB}=\sum_\alpha \lambda_\alpha A_\alpha\otimes B_\alpha,
\end{eqnarray}
with $A_\alpha$ being a Hermitian system operator to which the environment connects through the Hermitian bath operator $B_\alpha$ with dimensionless coupling $\lambda_\alpha\ll1$. We choose the system operator to be dimensionless with $||A_\alpha||\sim1$ and shift the energy dependence to the bath operator altogether. As a result, the characteristic coupling energies between system and bath would be controlled by $\lambda_\alpha$ making sure the weak coupling assumption is justified. Hereafter, we also incorporate $\lambda_\alpha$ into the bath operator $B_\alpha$.

As we are interested in the influence of the environment on the dynamics of the reduced system, first we need a qualitative description for the time evolution of the closed system, which means we possess a solution for the propagator in the Schr\"odinger equation $(\hbar=1)$,
\begin{eqnarray}
i\frac{\partial}{\partial t}U(t,t_0)=(H_0+V(t))U(t,t_0),\quad U(t_0,t_0)=\mathbb 1,
\end{eqnarray}
where we define $t_0$ to be the initial time instant when the driving is turned on, hence $V(t<t_0)=0$. The propagator evolves the state of the closed system, described by a density matrix from the initial time $t_0$ to a later time $t$; in the Schr\"odinger picture this is $\rho_S(t)=U(t,t_0)\rho_S(t_0)U^\dagger(t,t_0)$.

For later convenience, we introduce the interaction frame form of the unitary propagator,
\begin{eqnarray}
U(t,t_0)=e^{-iH_0t}\tilde U(t,t_0)e^{iH_0t_0},
\end{eqnarray} 
which obeys the equation of motion,
\begin{eqnarray}
\label{UI}
i\frac{\partial}{\partial t}\tilde U(t,t_0)=\tilde V(t)\tilde U(t,t_0),\quad \tilde U(t_0,t_0)=\mathbb 1.
\end{eqnarray}
Here, the interaction frame form of the driving term is $\tilde V(t)=e^{iH_0t}V(t)e^{-iH_0t}$. The state in the interaction frame is written as $\tilde \rho(t)=e^{iH_0t}\rho_S(t)e^{-iH_0t}$. We note that in standard time-dependent perturbation theory, this interaction frame would be called the interaction picture, however in the following we define the interaction picture with respect to the system-bath interaction, hence to avoid confusion we refer to the tilde notation as the interaction frame. 

Next, we define the interaction picture for the combined system-bath state and interaction,
\begin{eqnarray}
\label{Rint}
\rho_I(t)=U^\dagger(t,t_r)e^{iH_B(t-t_r)}\rho(t) e^{-iH_B(t-t_r)}U(t,t_r),\\
\label{Hint}
H_I(t)=U^\dagger(t,t_r)e^{iH_B(t-t_r)}H_{SB} e^{-iH_B(t-t_r)}U(t,t_r).
\end{eqnarray}
Here $t_r$ is an arbitrary reference time, when the Schr\"odinger picture and the interaction picture coincide. We can choose this reference time to be at the initial time $t_0$ when the drive is switched on. The entire time dependence is shifted to the interaction term and the von Neumann equation, $\dot\rho=-i[H,\rho]$, which describes the evolution of the combined state, becomes
\begin{gather}
\label{NEU}
\frac{\textrm d\rho_{I}}{\textrm dt}=-i[H_I(t),\rho_{I}(t)].
\end{gather}
We obtain the state of the reduced system by tracing out the environment,
\begin{gather}
\rho_{S}(t)=\textrm{Tr}_B\rho(t),\quad \rho_{I,S}(t)=\textrm{Tr}_B\rho_I(t),
\label{RSint}
\end{gather}
where the first is the reduced system state within the Schrödinger picture, and the second is the interaction picture reduced density operator. The states in the two pictures are connected by
\begin{gather}
\label{reduced}
\rho_{I,S}(t)=U^\dagger(t,t_0)\rho_{S}(t) U(t,t_0).
\end{gather}
It is crucial to emphasize that the states in Eq.~\eqref{RSint} are defined within the Hilbert space of $H_0$ which can encompass states beyond the computational subspace, so that the dynamics incorporate leakage effects. We further mention, that the interaction picture reduced density operator is not the same as the interaction frame density operator defined earlier, their relationship is given by
\begin{gather}
\rho_{I,S}(t)=e^{-iH_0t_0}\tilde U^\dagger(t,t_0)\tilde\rho(t) \tilde U(t,t_0)e^{iH_0t_0}.
\end{gather}

\subsection{The time-convolutionless master equation for driven quantum systems}

To determine the full dynamics, one has to solve the von Neumann equation \eqref{NEU}. However, our focus is exclusively on the dynamics of the reduced system, without consideration for the behavior of the bath degrees of freedom. By utilizing the time-convolutionless (TCL) projection operator technique (see Refs.~\cite{chaturvedi1979time,breuer,breuer2001time} for particular details of this method) to Eq. \eqref{NEU} we can describe the evolution of the interaction picture state of the reduced system by a time-local master equation, 
\begin{eqnarray}
\label{TCL1}
\frac{\textrm d}{\textrm dt}\rho_{I,S}(t)=-\int_{t_0}^t\mathrm ds\ \textrm{Tr}_B\left[H_I(t),[H_I(s),\rho_{I,S}(t)\otimes\rho_B]\right].\nonumber\\
\end{eqnarray}
The derivation of Eq.~\eqref{TCL1} relies on two explicit assumptions and one implicit assumption. First of all, factorizing initial system-bath state: $\rho(t_0)=\rho_{I,S}(t_0)\otimes\rho_B$, where $\rho_B$ is a time-independent reference state of the environment, which is chosen as an equilibrium state, $[H_B,\rho_B]=0$. 
Secondly, the weak coupling assumption: the exact time-local generator of the master equation is expanded up to the first nonvanishing order, which corresponds to the Born approximation.
These explicit assumptions are well justified for systems of interest for quantum information processing, as they are required by DiVincenzo's criteria \cite{crit}. The factorizing initial condition corresponds to the situation where the state of the system is initialized with a high fidelity and the weak coupling assumption yields long coherence times. Finally, the existence of the time-convolutionless generator implicitly presupposes the invertibility of the dynamical map. This imposes a self-consistency requirement on the solution of the master equation, mandating that it should always possess an inverse \cite{PhysRevA.101.012103}.  

We substitute Eqs.~\eqref{Rint}-\eqref{Hint} into Eq.~\eqref{TCL1} and subsequently transform the master equation into the Schr\"odinger picture and then into the interaction frame defined earlier, which yields,
\begin{eqnarray}
\label{ME}
\frac{\textrm d}{\textrm dt}\tilde\rho(t)=-i\left[\tilde V(t)+\tilde V_{\textrm{ren}}(t),\tilde\rho(t)\right]+\sum_{\alpha}\tilde D_\alpha(\tilde\rho).
\end{eqnarray}
The different decay channels are indexed by $\alpha$ as in Eq.~\eqref{eq:int}. The obtained master equation comprises a unitary component containing the drive term $\tilde V(t)$ as well as its renormalization $\tilde V_{\textrm{ren}}(t)$ due to the environmental effects. The remaining terms are dissipators, which effectively characterize the decaying influence of the environment on the system, 
\begin{align}
\tilde D_\alpha=\tilde A^{(f)}_\alpha(t)\tilde\rho_{}(t)\tilde A^\dagger_\alpha(t)-\frac{1}{2}\left\lbrace\tilde A^\dagger_\alpha(t) \tilde A^{(f)}_\alpha(t),\tilde\rho_{}(t)\right\rbrace+\textrm{ h.c.},\nonumber\\
\tilde V_{\textrm{ren}}=\frac{1}{2i}\sum_\alpha\left(\tilde A^\dagger_\alpha(t) \tilde A^{(f)}_\alpha(t)- \tilde A^{(f)\dagger}_\alpha(t)\tilde A_\alpha(t)\right)\label{diss}.
\end{align}
Each dissipator and renormalization term consists of a jump operator $\tilde A^{\dagger}_\alpha(t)$ associated with the respective system operator in the interaction Hamiltonian Eq.~\eqref{eq:int}, along with a filtered version,
\begin{eqnarray}
\label{filter1}
\tilde A^{(f)}_\alpha(t)=\sum_{\beta}\int_{t_0}^t\mathrm ds\ C_{\alpha\beta}(t,s)\tilde U(t)\tilde U^\dagger(s)\tilde A_\beta(s) \tilde U(s)\tilde U^\dagger(t).\nonumber\\ 
\end{eqnarray}
Here, the unitaries are the solutions of Eq.~\eqref{UI}, and to ease notation the dependence on the initial time $t_0$ is suppressed. 
The interpretation of the filtered operator is straightforward: the manipulation of the reduced system produces an input  $\tilde U^\dagger(s)\tilde A_\beta(s) \tilde U(s)$ for the convolution integral in Eq.~\eqref{filter1}, resulting in an environment response manifested in the time-dependence of $\tilde A_\alpha^{(f)}(t)$ with the correlation function $C_{\alpha\beta}(t,s)=\textrm{Tr}_B\left(B_\alpha(t)B_\beta(s)\rho_B\right)=C_{\alpha\beta}(t-s)$ acting as a response function. This filtered jump operator retains information from the past history of the dynamics, constituting the non-Markovian aspect of the master equation. 

In general, calculating the filtered jump operator and using it to construct the TCL dissipators and renormalization in Eq.~\eqref{diss} is very involved and typically requires demanding numerical computation. After the characterization of the driven quantum system by the definition of $H_0$ and the control protocol $V(t)$, one needs to solve Eq.~\eqref{UI}. The next crucial step is the identification of the environment and its coupling to the system, hence obtaining the correlation functions $C_{\alpha\beta}(t)$ and system operators  $A_\alpha$. At this point, one needs to evaluate multiple convolution integrals and finally solve the master equation to uncover the dynamics. It must be noted that the availability of this necessary information should be inherent, as the details of particular systems and environments are continuously under extensive research.

To enhance our understanding of the form of the dissipator and the decoherence it causes, we must choose a basis to construct the matrices of the filtered jump operators. 
Here, we pursue a description in the eigenbasis of the undriven reduced system as these are typically the states used to encode quantum information and we wish to see how decoherence acts on them. Using $H_0|n\rangle=E_n|n\rangle$, we write
\begin{eqnarray}
\label{frameA}
\tilde A_\alpha(t)=e^{iH_0t}A_\alpha e^{-iH_0t}=\sum_{\textbf{n}} e^{i\omega_{\textbf{n} } t}A_{\alpha,\textbf{n}}\hat P_\textbf{n} 
\end{eqnarray}
where the summation goes over the eigenstates labeled by $\textbf{n}\equiv\{n,m\}$, with Bohr frequencies $\omega_{\textbf{n}}=E_n-E_m$. We also defined the transition matrix element $A_{\alpha,\textbf{n}}=\langle n|A_\alpha|m\rangle $ and the matrix $\hat P_\textbf{n} =|n\rangle\langle m|$. 
By changing the integration variable in Eq.~\eqref{filter1} to $\tau=t-s$ and introducing the elapsed time $\Delta t=t-t_0$, the filtered jump operator becomes 

\begin{eqnarray}
\label{filter}
\tilde A^{(f)}_\alpha(t)=\sum_{\beta,\textbf{n}}e^{i\omega_{\textbf{n}} t}A_{\beta,\textbf{n}}F_{\alpha\beta,t}(\hat P_\textbf{n}),
\end{eqnarray}
where we defined the filtering operation, that transforms the matrix $\hat P_\textbf{n}$ by
\begin{align}
    F_{\alpha\beta,t}(\hat P_\textbf{n})=\int_{0}^{\Delta t}\mathrm d\tau\ C_{\alpha\beta}(\tau)e^{-i\omega_{\textbf{n}}\tau}\times\nonumber\\\tilde U(t,t-\tau)\hat P_\textbf{n}\tilde U^\dagger(t,t-\tau).
    \label{filtering}
\end{align}
Here, we have used the property $\tilde U(t,t_0)\tilde U^\dagger(t-\tau,t_0)=\tilde U(t,t_0)\tilde U(t_0,t-\tau)=\tilde U(t,t-\tau)$.  Combining all of the above expressions results in the following dissipator for the master equation of driven quantum systems, 
\begin{widetext}
\begin{equation}
\tilde D_\alpha(\tilde\rho)=\sum_{\beta,\textbf{n},\textbf{m}}e^{i(\omega_{\textbf{n}}-\omega_{\textbf{m}})t}A_{\beta,\textbf n}A^*_{\alpha,\textbf{m}}\biggl( \hat P_\textbf{n}\tilde\rho(t) F^\dagger_{\beta\alpha,t}(\hat P_\textbf{m})+
 F_{\alpha\beta,t}(\hat P_\textbf{n})\tilde\rho(t)\hat P^\dagger_\textbf{m}
-\frac{1}{2}\left\lbrace F^\dagger_{\beta\alpha,t}(\hat P_\textbf{m})\hat P_\textbf{n}+\hat P^\dagger_\textbf{m}F_{\alpha\beta,t}(\hat P_\textbf{n}),\tilde\rho(t)\right\rbrace\biggr).
\label{dissipator1}
\end{equation}
\end{widetext}
The evaluation of the filtering operation is the most important component in this dissipator, as it determines the jump processes and their time-dependent rates. In the next subsection, we review previously made approximations in the calculation of the filtering operation. 

\subsection{The field-independent and adiabatic approximations}
In the absence of any time-dependence in the Hamiltonian \eqref{eq:ham}, i.e. $V(t)=0$, the solution of \eqref{UI} yields $\tilde U(t,t-\tau)=\mathbb 1$ and the filtering operation becomes
\begin{eqnarray}
\label{eq:19}
F_{\alpha\beta,t}(\hat P_\textbf{n})=\Gamma_{\alpha\beta}(\omega_{\textbf{n}},\Delta t)\hat P_\textbf{n},
\end{eqnarray}
which is just the multiplication of the matrix $\hat P_\textbf{n}$ with the time-dependent rate 
\begin{eqnarray}
\label{rate0}
\Gamma_{\alpha\beta}(\omega_{\textbf n},\Delta t)=\int_{0}^{\Delta t}\mathrm d\tau\ C_{\alpha\beta}(\tau)e^{-i\omega_{\textbf n}\tau}.
\end{eqnarray}
In this case, the form of the dissipator in Eq.~\eqref{dissipator1} leads to the Redfield master equation in Eq.~\eqref{ME} and with a subsequent Markovian approximation with the elapsed time taken to infinity ($\Delta t\to\infty$), we obtain the Bloch-Redfield theory \cite{blocholder,redfield}. The well-known Redfield dissipator reads,
\begin{gather}
\tilde D^{(R)}_\alpha=\sum_{\beta,\textbf n,\textbf m}e^{i(\omega_{\textbf n}-\omega_{\textbf m})t}\left(\Gamma_{\alpha\beta}(\omega_{\textbf n},\Delta t)+\Gamma^*_{\beta\alpha}(\omega_{\textbf m},\Delta t)\right)\times\nonumber\\
A_{\beta,\textbf n}A^*_{\alpha,\textbf m}\left(\hat P_\textbf{n}\tilde\rho(t) \hat P^\dagger_\textbf{m}-\frac{1}{2}\left\lbrace \hat P^\dagger_\textbf{m}\hat P_\textbf{n},\tilde\rho(t)\right\rbrace\right).\label{redfield}
\end{gather}
This dissipator resembles the generalized Lindblad form; however, its decoherence matrix is not guaranteed to be positive definite. As a result, it still leads to a non-Markovian master equation. By employing further approximations, such as time coarse-graining \cite{coarse1} or its extreme version \cite{Whitney_2008}, the secular approximation, one can transform the dissipator into the Lindblad form and obtain a Markovian master equation. These approximations come with the benefit of avoiding complications associated with the resulting dynamics losing their complete positivity \cite{pos1,Davidovic2020completelypositive}, something that may occur with the TCL master equation. On the other hand, it has been shown that, e.g., the secular approximation may lead to results that are physically not justified \cite{secular,secular2}. Here, we avoid these approximations altogether.

At this point, we mention that even in the presence of a driving term in the Hamiltonian, one often unknowingly makes the approximation, 
\begin{eqnarray}
\label{uncorrapprox}
\tilde U(t,t-\tau)\approx\mathbb 1,
\end{eqnarray}
in Eq.~\eqref{filtering}. The result is the field-independent master equation in which the drive term only appears in the Hamiltonian component and the terms responsible for the decaying dynamics are completely unaffected by the control of the system. 
However, as mentioned in the introduction previously, this approximation implies a drastic simplification of the effects of quantum noise acting on driven systems. The assumption is that decoherence acts independently from the controlled self-evolution of the closed system, and the decaying effects of the environment can only propagate through the dynamics \cite{nonloc}. 

The field-independent approximation is valid for delta-correlated environments, $C_{\alpha\beta}(\tau)\sim\delta(\tau)$, which is an idealization. A more realistic approach takes into account the finite but short decay time of the correlation functions. In this case, the Markov approximation facilitates the adiabatic approximation of the unitary,
\begin{eqnarray}
\label{adiabaticapprox}
\tilde U(t,t-\tau)\approx\exp\left(-i\tilde V(t)\tau\right),
\end{eqnarray}
leading to the adiabatic quantum master equations.
Here, we wish to move beyond these approximations.

\subsection{Beyond the zeroth of the Magnus expansion}\label{subs:mag}

In the following, we restrict the discussion to driven finite-dimensional quantum systems. 
To evaluate the filtering operation in Eq.~\eqref{filtering}, we first need to calculate a matrix product and then a convolution integral. Due to the weak-coupling assumption, we do not need to use the exact solution for the unitary $\tilde U$, and it suffices to use an approximation that captures the qualitative properties of the closed dynamics. We exploit this idea and express the unitary based on its Magnus expansion \cite{magnus,magnus1},
\begin{eqnarray}
\label{formalexact}
\tilde U(t,t-\tau)=\exp\left(-i\int_{t-\tau}^{t}\textrm dt_1\ \tilde V_{\textrm{eff}}(t_1,\tau)\right),
\end{eqnarray}
where the effective interaction frame drive Hamiltonian can be constructed by the recurrence relation of nested commutators in the Magnus series, see Appendix~\ref{magna} for further details. Up to the second order, the effective Hamiltonian reads,
\begin{eqnarray}
\label{veff}
\tilde V_{\textrm{eff}}(t,\tau)\approx\tilde V(t)-\frac{i}{2}\int_{t-\tau}^t\textrm dt_1\ [\tilde V(t),\tilde V(t_1)].
\end{eqnarray}
This effective Hamiltonian description of the dynamics has a long history in the literature of nuclear magnetic resonance and it is usually referred to as average Hamiltonian theory \cite{NMR2,NMR1,avgham}. We remark that the zeroth order corresponds to $\tilde V_{\textrm{eff}}(t,\tau)\approx0$ which yields Eq.~\eqref{uncorrapprox} while the first order $\tilde V_{\textrm{eff}}(t,\tau)\approx\tilde V(t)$ combined with the Markovian approximation leads to Eq.~\eqref{adiabaticapprox}.


Without loss of generality, we write the time-dependent control term $V(t)$ in the eigenbasis of $H_0$ as a traceless Hermitian matrix. Consequently, the matrix of $\tilde V(t)$ may be written as a linear combination of generalized Gell-Mann matrices, the higher-dimensional extensions of the Pauli matrices, the standard SU$(N)$ generators \cite{GGM,sun}. They are defined as three different types of matrices, written in the standard basis,
\begin{eqnarray}
    \Lambda_{jk}^s=|j\rangle\langle k|+|k\rangle\langle j|,\ 0\leq j<k\leq N-1,\\
\Lambda_{jk}^a=-i(|j\rangle\langle k|-|k\rangle\langle j|),\ 0\leq j<k\leq N-1,\\
\Lambda_{l}^d=\sqrt{\frac{2}{l(l+1)}}\left(\sum_{j=0}^{l-1}|j\rangle\langle j|-l|l\rangle\langle l|\right),
\label{gelldiag}
\end{eqnarray}
where $1\leq l\leq N-1$. There are $N(N-1)/2$ symmetric and anti-symmetric generalized Gell-Mann matrices each, and $N-1$ diagonal ones. When gathered into the vector $\mathbf\Lambda$, we order the components as symmetric, anti-symmetric, and diagonals.
The generalized Gell-Mann matrices form a closed group with respect to matrix multiplication. Since the effective Hamiltonian is given by a sum of nested commutators, the unitary in Eq.~\eqref{formalexact} must be of the form
\begin{eqnarray}
\label{unitarymatrix1}
\tilde U(t,t-\tau)=\exp\left(-i\mathbf \Lambda\cdot\mathbf r(t,\tau)\right),
\end{eqnarray}
where $\mathbf r$ is an $N^2-1$ dimensional vector whose components $r_k(t,\tau)$ are obtained from the Magnus expansion. We may now expand the matrix exponential in Eq.~\eqref{unitarymatrix1} into the generalized Gell-Mann matrices and the unit matrix \cite{sun}.
To achieve this, we define the eigenvalues of $\mathbf\Lambda\cdot\mathbf r$ as $\mu_j(\mathbf r),j=1,...,N$, where the time-dependence is suppressed to ease notations ($\mathbf r\equiv\mathbf r(t,\tau)$). We further define the characteristic function
\begin{eqnarray}
\label{char}
K(\mathbf r)=\sum_{j=1}^N e^{-i\mu_j(\mathbf r)},
\end{eqnarray}
which enables us to write
\begin{eqnarray}
\label{unitarymatrix2}
\tilde U(t,t-\tau)=\frac{K(\mathbf r)}{N}\mathbb 1+\frac{i}{2}\nabla K(\mathbf r)\cdot\mathbf\Lambda,
\end{eqnarray}
with $\nabla_k=\partial/\partial r_k\equiv \partial_k$. The characteristic function and its derivatives completely determine the unitary. Furthermore, the characteristic function can be seen as a renormalization tool for the time-dependent rates in Eq.~\eqref{rate0} that would appear in the Redfield part of the dissipator. Based on Eq.~\eqref{unitarymatrix2}, the filtering operation in Eq.~\eqref{filtering} can be written as
\begin{eqnarray}
\label{filtering2}
F_{\alpha\beta,t}(\hat P_\textbf{n})=\tilde\Gamma_{\alpha\beta}(\omega_{\textbf n}, t)\hat P_\textbf{n}+\hat M_{\alpha\beta}(\omega_\textbf n,t),
\end{eqnarray}
where the first term is the renormalization of the rates appearing Eq.~\eqref{eq:19} and it is obtained using only the diagonal part of Eq.~\eqref{unitarymatrix2}. The second term is a correction matrix, that does not contain any element proportional to $\hat P_\textbf{n}$. The detailed formulas for the renormalization and correction matrix are found in Appendix~\ref{hugeformulas}. Utilizing Eq.~\eqref{filtering2}, we can write the dissipator for driven systems in the standard basis of $H_0$ as
\begin{eqnarray}
\tilde D_\alpha(\tilde\rho)=\tilde D^{(R)}_\alpha(\tilde\rho)+\tilde D^{(C)}_\alpha(\tilde\rho),  
\end{eqnarray}
where the first term is the Redfield dissipator of Eq.~\eqref{redfield} in which the standard rates valid for undriven systems are renormalized, $\Gamma_{\alpha\beta}\to\tilde \Gamma_{\alpha\beta}$, to account for the presence of the drive, whereas the second term is the correction to the Redfield form due to the presence of the drive. This correction is described by Eq.~\eqref{dissipator1} by the substitution $F_{\alpha\beta,t}\to \hat M_{\alpha\beta}$. We remark that the characterization of the renormalization Hamiltonian $\tilde V_{\textrm{ren}}$ is straightforward. The field-independent approximation is obtained for $K(\mathbf r)\approx N$.


\section{\label{sec:level3}General driven qubit}

Here, we illustrate the derivation of the TCL dissipator for a generally driven two-level system. We denote the ground state by $|0\rangle$ with corresponding energy $E_0$, and the excited state by $|1\rangle$ and energy $E_1$, resulting in the qubit frequency  $\omega_q=E_1-E_0$. An external time-dependent field induces a coupling between these states. Without loss of generality, we take this field to be real and write the Hamiltonian of the system as
\begin{eqnarray}
    H(t)=-\frac{\omega_q}{2}\sigma_z+\Omega(t)\sigma_x.
\end{eqnarray}
Here, we defined the zero of energy halfway between $E_0$ and $E_1$. The Pauli matrices emerge as $\sigma_z=|0\rangle\langle0|-|1\rangle\langle1|$ and $\sigma_x=|0\rangle\langle1|+|1\rangle\langle0|$. To characterize the closed dynamics of the reduced system qualitatively, we use the Magnus expansion up to second order. We expand the interaction frame effective Hamiltonian in Eq.~\eqref{veff} with the Pauli matrices and obtain
\begin{eqnarray}
\label{qubiteff}
    \tilde V_{\textrm{eff}}(t,\tau)=\Omega(t)\cos(\omega_qt)\sigma_x+\Omega(t)\sin(\omega_qt)\sigma_y\nonumber\\
    -\Omega(t)\int_{t-\tau}^t\textrm dt_1\ \Omega(t_1)\sin(\omega_q(t-t_1))\sigma_z.
\end{eqnarray}
The higher-order odd terms in the Magnus series correct the coefficients of $\sigma_{x,y}$ while the even terms contribute to the coefficient of $\sigma_z$. Integrating Eq.~\eqref{qubiteff} from $t-\tau$ to $t$ yields the vector $\mathbf r(t,\tau)$ in Eq.~\eqref{unitarymatrix1}. The characteristic function Eq.~\eqref{char} for the driven qubit is simply
\begin{eqnarray}
    K(\mathbf r)=2\cos r,
\end{eqnarray}
with $r=|\mathbf r|$. Using Eq.~\eqref{unitarymatrix2}, the unitary that appears in the filtering operation is then
\begin{eqnarray}
\label{qubitu}
   \tilde U(t,t-\tau)=\cos r\mathbb 1-i\frac{\sin r}{r}\mathbf r\cdot\bm\sigma,
\end{eqnarray}
as it should be for an exponential involving Pauli matrices.

After the characterization of the closed system dynamics, we turn our attention to the coupling between the system and the bath. The most general two-level system operator to which the environment can couple is of the form, $a_x\sigma_x+a_y\sigma_y+a_z\sigma_z$, where $a_{x,y,z}\in\mathbb{R}$. To simplify the coming examples, we first consider only a longitudinal coupling to the environment, and later a transverse coupling along only one axis. These are very important examples as they describe pure dephasing and relaxation, respectively. Considering the properties of the environment, the only relevant microscopic detail is the correlation function involving the bath observables appearing in the system-bath interaction Hamiltonian. We further use the noise power spectral density, which is defined as the Fourier transform of the correlation function,
\begin{eqnarray}
\label{NP}
    S(\omega)=\int_{-\infty}^\infty \textrm dt\ C(t)e^{i\omega t}\leftrightarrow\ C(t)=\int_{-\infty}^\infty \frac{\textrm d\omega}{2\pi}\  S(\omega)e^{-i\omega t}.\nonumber\\
\end{eqnarray}
By defining the symmetrized noise power spectral density $2\bar S(\omega)=S(\omega)+S(-\omega)$ and the antisymmetrized noise power spectral density $2J(\omega)=S(\omega)-S(-\omega)$, we may write the real and imaginary part of the correlation function as,
\begin{eqnarray}
\label{RNP}
    \textrm{Re}(C(t))=\int_{-\infty}^\infty \frac{\textrm d\omega}{2\pi}\  \bar S(\omega)e^{i\omega t},\\
    \label{INP}
    \textrm{Im}(C(t))=-\int_{-\infty}^\infty \frac{\textrm d\omega}{2\pi i}\  J(\omega)e^{i\omega t}.
\end{eqnarray}

\subsection{Qubit dephasing}
A bath coupled longitudinally to the qubit causes fluctuations in its frequency which leads to dephasing. As such, the system operator appearing in $H_{\textrm{SB}}$ is $A_\varphi=\sigma_z$. Moving into the interaction frame, we have $\tilde\sigma_z(t)=\sigma_z$, hence we only need to evaluate the filtering operation on one matrix. Substituting Eq.~\eqref{qubitu} to Eq.~\eqref{filtering} reveals the filtering $F_{\varphi,t}(\sigma_z)$ is a linear combination of Pauli matrices and we may write
the dephasing dissipator for a driven qubit as
\begin{eqnarray}
\label{Dphi}
    \tilde D_\varphi(\tilde \rho)=\gamma_\varphi(t)\left(\sigma_z\tilde\rho(t)\sigma_z-\tilde\rho(t)\right)+\tilde D^{(C)}_\varphi(\tilde \rho).
\end{eqnarray}
Here, the first term is the standard pure dephasing dissipator with a time-dependent rate
\begin{eqnarray}
\label{deph1}
    \gamma_\varphi(t)=2\int_0^{\Delta t}\textrm d\tau\ \textrm{Re}(C_\varphi(\tau))\left(1-2\frac{r_x^2+r_y^2}{r^2}\sin^2r\right).\nonumber\\
\end{eqnarray}
Interestingly, this is exactly the dephasing rate valid for the undriven system minus corrections. As these correction terms are all positive, driving the system decreases the dephasing rate provided the real part of the correlation function is also positive. The price to pay for this decrease is the opening of additional decay channels, which is described by
the second term in Eq.~\eqref{Dphi},
\begin{eqnarray}
    \tilde D^{(C)}_\varphi(\tilde \rho)=\gamma_x(t)\left(\sigma_x\tilde\rho(t)\sigma_z-\frac{1}{2}\{\sigma_z\sigma_x,\tilde\rho(t)\}\right)\nonumber\\
    +\gamma_y(t)\left(\sigma_y\tilde\rho(t)\sigma_z-\frac{1}{2}\{\sigma_z\sigma_y,\tilde\rho(t)\}\right)+\textrm{h.c.},
\end{eqnarray}
with rates
\begin{eqnarray}
    \gamma_x(t)=\int_0^{\Delta t}\textrm d\tau\ C_\varphi(\tau)\left(\frac{r_y}{r}\sin2r+2\frac{r_xr_z}{r^2}\sin^2r\right),\nonumber\\ \\
    \gamma_y(t)=\int_0^{\Delta t}\textrm d\tau\ C_\varphi(\tau)\left(-\frac{r_x}{r}\sin2r+2\frac{r_yr_z}{r^2}\sin^2r\right).\nonumber\\
\end{eqnarray}
As there is only one decay channel in the field-independent approach, the appearance of the additional channels changes the physics significantly, e.g., the predicted asymptotic states by the two approaches are strikingly different, see Fig.~\ref{Rabifig}.

For the driven qubit, the renormalization of the Hamiltonian in case of longitudinal coupling to the environment can be simply written as,
\begin{eqnarray}
\label{phiren}
\tilde V_{\textrm{ren},\varphi}(t)=\textrm{Re}(\gamma_x(t))\sigma_y-\textrm{Re}(\gamma_y(t))\sigma_x,
\end{eqnarray}
which clearly describes a drive renormalization. 

\subsection{\label{rabi}Rabi driving and pure dephasing}
To illustrate the effects of the renormalization terms and the correction dissipator, we discuss the simple example of a Rabi-driven two-level system in the presence of pure dephasing noise. In this case, the driving is simply written as $\Omega(t)=\Omega_x\cos\omega_dt+\Omega_y\sin\omega_dt$, henceforth we assume resonant driving $\omega_d=\omega_q$. Within the rotating wave approximation (RWA), we drop every fast-rotating terms. As a result, the vector $\mathbf r$ greatly simplifies and reads
\begin{eqnarray}
    r_x(t,\tau)=\frac{\Omega_x\tau}{2},\quad r_y(t,\tau)=\frac{\Omega_y\tau}{2},\quad r_z(t,\tau)=0.
\end{eqnarray}
The higher-order Magnus terms all vanish within the RWA. The closed dynamics is described by rotations around the Bloch sphere, for $\Omega_y=0$ the rotation axis is $x$, and for $\Omega_x=0$ it is the $y$-axis, if both are non-zero the axis lies somewhere in the $xy$ plane. Starting the dynamics initially in the ground state, the drive first rotates the state to the excited state and subsequently back to the ground state giving rise to Rabi cycles.

We can write the dephasing rate in Eq.~\eqref{deph1} as
\begin{eqnarray}
\label{eq:dephasing_rate}   \gamma_\varphi(t)=2\int_0^{\Delta t}\textrm d\tau\ \textrm{Re}(C_\varphi(\tau))\cos(\Omega_R\tau),
\end{eqnarray}
where we defined the Rabi frequency $\Omega_R=\sqrt{\Omega_x^2+\Omega_y^2}$. For elapsed times $\Delta t$ much longer than the time scale over which the real part of the correlation function $C_\varphi(\tau)$ vanishes, the rate approaches the asymptotic value,
\begin{eqnarray}
    \gamma_\varphi(\infty)=\bar S_\varphi(\Omega_R),
\end{eqnarray}
where we substituted Eq.~\eqref{RNP} into Eq.~\eqref{eq:dephasing_rate}. 
This is in great contrast to the field-independent description, where the dephasing rate approaches the value $\bar S_\varphi(0)$ as is well-known from the standard theory of weakly coupled dephasing noise \cite{drivenspin}. The correction due to the drive on the dephasing rate is at least first order in the Rabi frequency, $\gamma_\varphi\approx\bar S_\varphi(0)+\Omega_R\bar S_\varphi'(0)$.
\begin{figure*}
    \centering
    \includegraphics[width=1\linewidth]{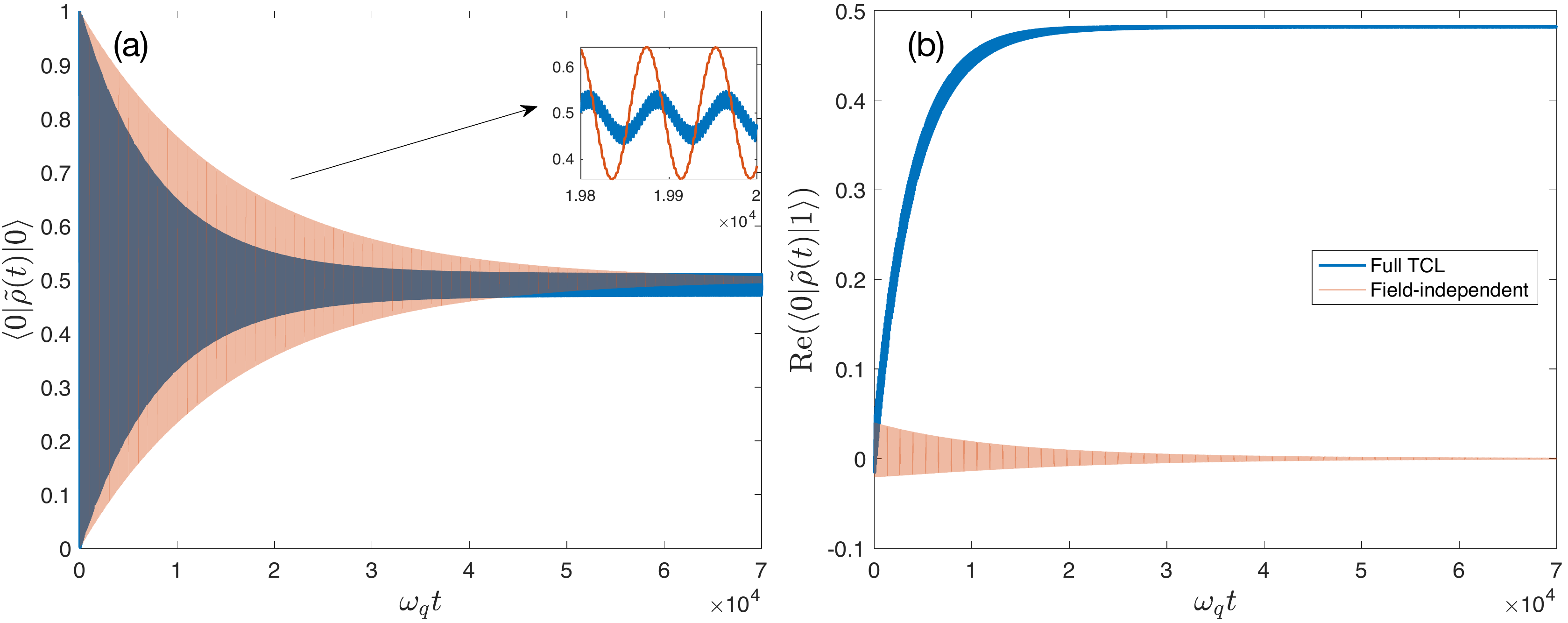}
    \caption{Both figures depict a Rabi-driven two-level system in the presence of dephasing noise, modeled as an Ohmic boson bath with symmetrized noise power spectral density $\bar S_\varphi(\omega)=\lambda\omega e^{-\omega/\omega_c}\coth(\beta\omega/2)$. (a) Decaying Rabi oscillations of the population of the state $|0\rangle$. The red curve corresponds to the dynamics within the field-independent approximation, while the blue curve is obtained by keeping all contributions in the TCL master equation. The timescales over which the Rabi oscillations decay are visibly different as the asymptotic decay rates corresponding to the two different approaches are $\bar S_\varphi(0)<\bar S_\varphi(\Omega_R)$. The inset shows the shift in the frequency of Rabi oscillations. (b) The real part of the off-diagonal element.   The density matrix asymptotically obtains significantly different off-diagonal values in the two cases. The non-equilibrium steady state corresponds to the mixed state of the states $|\pm\rangle$ at inverse temperature $\beta$ in the full TCL approach, while it is the completely mixed state of the states $|0\rangle$ and $|1\rangle$ in the field-free approach.  The parameters used, Rabi frequency: $\Omega_R=0.08\omega_q$, inverse temperature: $\beta\omega_q=50$, high frequency cutoff: $\omega_c=5\omega_q$, dimensionless system-bath coupling: $\lambda=10^{-4}$.}
    \label{Rabifig}
\end{figure*}

The additional rates appearing in the correction dissipator and renormalization Hamiltonian may be written as
\begin{eqnarray}
    \gamma_{x,y}(t)=\pm\frac{\Omega_{y,x}}{\Omega_R}\int_0^{\Delta t}\textrm d\tau\ C_\varphi(\tau)\sin(\Omega_R\tau).
\end{eqnarray}
These rates asymptotically approach the values
\begin{eqnarray}
    \gamma_{x,y}(\infty)=\mp\frac{\Omega_{y,x}}{\Omega_R}\left(\frac{iJ_\varphi(\Omega_R)}{2}+\dashint_{-\infty}^\infty\frac{\textrm d\omega}{2\pi}\  \frac{\bar S_\varphi(\omega)\Omega_R}{\omega^2-\Omega_R^2}\right),\nonumber\\
\end{eqnarray}
with the real part being a principal value integral. For asymptotic times the interaction-frame drive Hamiltonian is approximated with
\begin{eqnarray}
    \tilde V+\tilde V_{\textrm{ren},\varphi}\approx \left(\frac{\Omega_x}{2}-\textrm{Re}(\gamma_y)\right)\sigma_x+\left(\frac{\Omega_y}{2}+\textrm{Re}(\gamma_x)\right)\sigma_y,\nonumber\\
\end{eqnarray}
from which we see the real parts of the correction rates $\gamma_{x,y}$ eventually shift the frequency of the Rabi oscillations. 

In the simple case when $\Omega_y=0$, the asymptotic solution of the master equation in the RWA is analytically available and reads
\begin{eqnarray}
    \tilde\rho(t\to\infty)=\frac{1}{2}\begin{pmatrix}
        1&\frac{J(\Omega_R)}{\bar S(\Omega_R)}\\
        \frac{J(\Omega_R)}{\bar S(\Omega_R)}&1
    \end{pmatrix}.
\end{eqnarray}
This is a mixed state of the states $|\pm\rangle$ with probabilities determined by the ratio of the antisymmetrized and symmetrized noise power spectrum. Hence, asymptotic coherence can be achieved in the presence of a quantum bath.
For example, for a thermal quantum environment, the ratio is $J(\Omega_R)/\bar S(\Omega_R)=\tanh(\beta\Omega_R/2)$ due to the fluctuation-dissipation theorem. We depict all of the above-mentioned effects on an example in Fig.~\ref{Rabifig}.

\subsection{Qubit relaxation}
The transverse coupling $A_R=\sigma_x$ allows energy exchange between the qubit and the environment leading to the relaxation of the qubit. In the interaction frame, the relevant system operator is $\tilde\sigma_x(t)=e^{i\omega_qt}\sigma_-+e^{-i\omega_qt}\sigma_+$, with $\sigma_-=|1\rangle\langle0|$ and $\sigma_+=|0\rangle\langle1|$. The filtering operation is carried out for the matrices $\sigma_\pm$, which leads to the following dissipator,
\begin{equation}
 \label{Drelax}
    \tilde D_R(\tilde \rho)=\sum_{i,j=\pm,z}\gamma_{ij}(t)\left(\sigma_i\tilde\rho(t)\sigma_j^\dagger-\frac{1}{2}\{\sigma^\dagger_j\sigma_i,\tilde\rho(t)\}\right).
 \end{equation}
Here, the dissipator has been cast into the generalized Lindblad form. 
There are two non-zero diagonal elements in the decoherence matrix $\gamma_{ij}(t)$, the renormalized time-dependent emission and absorption rates, while the third diagonal is $\gamma_{zz}=0$.
The off-diagonal elements for $\pm$ correspond to the non-secular terms and originate from the Redfield dissipator, while the off-diagonals for $+z$ and $-z$ are due to the correction dissipator.  We report the precise formulas of the rates appearing in the decoherence matrix in Appendix~\ref{relaxapp}. 

The decoherence matrix is Hermitian, its diagonalization $\gamma_{ij}(t)=\sum_kW_{ik}(t)d_k(t)W^*_{jk}(t)$, yields real eigenvalues $d_k$ which correspond to the canonical rates of the dissipator \cite{canon}. After defining the jump operators $L_k(t)=\sum_i W_{ik}(t)\sigma_i$, where $W_{ik}$ are the components of the eigenvectors of the decoherence matrix, the dissipator can be rewritten in the canonical Lindblad form $\tilde D_R(\tilde \rho)=\sum_kd_k(t)\left(L_k(t)\tilde\rho(t)L_k^\dagger(t)-1/2\{L_k^\dagger(t)L_k(t),\tilde\rho(t)\}\right)$, revealing the canonical decay channels. For a generally driven qubit that is coupled transversely to a quantum bath, the decoherence matrix always has a zero eigenvalue, as we show in Appendix~\ref{relaxapp}. Consequently, weak transverse coupling generally leads to two canonical decay channels for qubits characterized by the rates $d_k(t)$ and jump operators $L_k(t)$, with $k=1,2$. One of the canonical rates can take negative values for certain times, corresponding to non-Markovian evolution \cite{canon,eternal,prb,simple}. Within the field-independent approximation, these two channels are described by the Redfield dissipator, which captures the effects of these channels very well. We can contrast this with the case of longitudinal coupling to the environment which leads to the dissipator $\tilde D_\varphi$ in Eq.~\eqref{Dphi}. Again, $\tilde D_\varphi$ can be transformed into the canonical Lindblad form, revealing two canonical decay channels. There, the field-independent approximation fails to correctly characterize the channels, as it ignores one of them altogether. 

The Markovian approximation combined with the secular approximation leads to the standard phenomenological description of relaxation with the canonical rates becoming constants and the jump operators are approximated as $L_k(t)\approx\sigma_\pm$. 
Following the simple Rabi driving example from the previous subsection but in the presence of relaxation noise, the aforementioned constant canonical rates $d_k$ are approximated as the asymptotic values of the renormalized emission and absorption rates that appear as the first two elements in the diagonal of the decoherence matrix $\gamma_{ij}$; these are (see also the Appendix~\ref{relaxapp}),
\begin{gather}
    \gamma_{\pm}(\infty)=\int_{-\infty}^\infty \frac{\textrm d\omega}{2\pi}\  S(\omega)\int_{0}^\infty \textrm d\tau\ \cos(\omega-\omega_q)\tau\cos^2\frac{\Omega_R\tau}{2}\nonumber\\=\frac{S(\pm\omega_q)}{2}+\frac{1}{4}\left(S(\pm\omega_q+\Omega_R)+S(\pm\omega_q-\Omega_R)\right).
\end{gather}
For time-independent systems, the rates are obtained using Fermi's golden rule to yield $\gamma_\pm=S(\pm\omega_q)$. For weakly driven systems, i.e. when $\Omega_R\ll\omega_q$,
we may expand the noise power spectrum around $\pm \omega_q$ and obtain the corrections to the golden rule decay rates,
\begin{eqnarray}\label{relaxrate}
    \gamma_\pm(\infty)=S(\pm\omega_q)+\frac{\Omega_R^2}{4}S''(\pm\omega_q)+\mathcal O(\Omega_R^4),
\end{eqnarray}
with $S''$ denoting the second derivative of the noise spectrum. The correction to the relaxation rates are second order in the Rabi frequency, in contrast to dephasing noise where it was first order. 

\subsection{Possible consequences for the fidelities of single-qubit gates}

A very simple formula for the average fidelity of single-qubit operations in the presence of an amplitude damping channel, i.e. relaxation, may be written as,
\begin{eqnarray}
    \mathcal F(t_g) = \frac{1}{6}\left(3+e^{-\gamma t_g}+2e^{-\gamma t_g/2}\right),
\end{eqnarray}
with $t_g$ being the gate time and $\gamma$ the relaxation rate. We approximate the relaxation rate by $\gamma\approx\gamma_+(\infty)+\gamma_-(\infty)\approx\gamma_+(\infty)$ from Eq.~\eqref{relaxrate} as the absorption rate $\gamma_-(\infty)$ is usually suppressed at low temperatures at which most quantum processors are operated. For short gate times, the infidelity, which quantifies in this case the incoherent error of the gate operation \cite{PEDERSEN}, is approximated by
\begin{eqnarray}
\label{errorrelax}
    1-\mathcal F(t_g) \approx\frac{S(\omega_q) t_g}{3}+(\Omega_R t_g)^2\frac{S''(\omega_q)}{12t_g}.
\end{eqnarray}
Here, it is crucial to understand that the gate time and the Rabi frequency are not independent, their product determines the rotation angle around the Bloch sphere, hence $\Omega_Rt_g=\mathrm{const}$. The field-independent approximation retains only the first term and it predicts that decreasing the gate time may decrease the incoherent errors indefinitely. Equation \eqref{errorrelax} implies that it is impossible to outrun the decaying effects of the environment and that there is an optimal gate time due to the competition of the field-independent and field-dependent contributions \cite{alicki1}. 

A similar argument can be made for the incoherent errors caused by the dephasing channel \cite{PEDERSEN}, which is approximated by
\begin{eqnarray}\label{gatefid}
    1-\mathcal F_\varphi(t_g) \approx \frac{\gamma_\varphi t_g}{3}\approx\frac{\bar S_\varphi(0)t_g}{3}+ \Omega_Rt_g\frac{\bar S'_\varphi(0)}{3}.
\end{eqnarray}
The constant shift appearing in the last term of this error estimate was assumed in Ref.~\cite{fastgate} without understanding its origin to explain the discrepancy between measurement and simulation data obtained within the field-independent approximation.

\section{\label{sec:level4}The driven transmon and leakage}

The transmon qubit is based in superconducting technology and is created by capacitively shunting a Josephson tunneling junction or a superconducting quantum interference device \cite{transmon1}. These superconducting circuits realize an artificial atom that can be modeled as a weakly anharmonic oscillator. The control of such a system is achieved by microwave pulses driving resonantly the transition between the two lowest-lying levels \cite{scqubit1}. These levels span the computational subspace while transitions (``leakage") into all other oscillator states represent qubit errors. The Hamiltonian of the controlled transmon modeled as a Duffing oscillator can be written in the laboratory frame as $H=H_0+V(t)$ where,
\begin{eqnarray}
\label{duff}
H_0=\omega_q a^\dagger a+\frac{\Delta_a}{2}a^\dagger a^\dagger a a,\\
V(t)=\Omega(t)\left(a+a^\dagger\right).
\label{drive}
\end{eqnarray}
Here, $a^{(\dagger)}$ are the standard annihilation and creation operators of the harmonic oscillator, with $\omega_q$ the qubit frequency and $\Delta_a$ the anharmonicity. The eigenstates of $H_0$ are the standard harmonic oscillator states which we denote $|n\rangle$, $n\in\mathbb N$. The computational subspace is then $n=0,1$ and $n\geq2$ are the leakage levels. Equation \eqref{drive} describes the control of the transmon with the charge operator $a+a^\dagger$ and $\Omega(t)$ describing a voltage pulse, typically of the form,
\begin{eqnarray}
\label{transmondrive}
\Omega(t)=\Omega_x(t)\cos\omega_d t+\Omega_y(t)\sin\omega_d t.
\end{eqnarray}
In what follows, we assume resonant driving of the lowest-lying levels, i.e., $\omega_d=\omega_q$, and set the initial time to $t_0=0$.

On the one hand, similarly to the Rabi-driven two-level system in Sec.~\ref{rabi}, the voltage pulse in Eq.~\eqref{transmondrive} manipulates the state in the computational subspace and creates rotations around the Bloch sphere of the transmon qubit. For instance, creating a rotation around the $x$-axis by an angle $\theta$ requires,  $\int_0^{t_g}\textrm dt\  \Omega_x(t)=\theta$ where $t_g$ denotes the duration of the pulse and $\Omega_y(t)=0$. On the other hand, Eq.~\eqref{transmondrive} also acts as an off-resonant drive for the higher-lying levels of the anharmonic oscillator, populating these states and creating leakage. This can be avoided by pulse shaping, which is the careful tailoring of the envelopes $\Omega_{x,y}(t)$. The standard methods for superconducting qubits are Gaussian shaping and derivative removal (DRAG) \cite{drag1,drag2}, which entails the form of the envelopes for $x$-axis rotations to be
\begin{eqnarray}
    \Omega_x(t)&=&\theta\frac{\exp\left(-\frac{(t-t_g/2)^2}{2\sigma^2}\right)-\exp\left(-\frac{t_g^2}{8\sigma^2}\right)}{\sqrt{2\pi\sigma^2}\textrm{erf}(t_g/\sqrt{8\sigma^2})-t_g\exp\left(-\frac{t_g^2}{8\sigma^2}\right)},\\
    \Omega_y(t)&=&-\xi\frac{\dot\Omega_x(t)}{\Delta_a}.
    \label{dragy}
\end{eqnarray} 
The $x$ quadrature amplitude are chosen such that the pulse duration is $t_g$, implying $\Omega_x(0)=\Omega_x(t_g)=0$, and such that the correct amount of rotation is implemented $\int_0^{t_g}\textrm dt\  \Omega_x(t)=\theta$. The essence of Gaussian shaping lies in the fact that the frequency bandwidth of the pulse, denoted by $1/\sigma$ where $\sigma$ is the standard deviation, limits the spectral weight of the pulse at the transition frequency to higher levels. However, for short pulses, there is still significant spectral weight at $\Delta_a$ which increases leakage for example to the state $|2\rangle$. This is countered by the $y$ quadrature amplitude in Eq.~\eqref{dragy}, which acts as the DRAG correction where $\xi$ is an optimization parameter. Henceforward, we are interested in $X_{\pi/2}$ pulses, as any SU$(2)$ gate can be generated using only $X_{\pi/2}$ pulses combined with virtual Z gates \cite{zgate}.

\subsection{Qutrit approximation}
In the following, we quantify average leakage and gate fidelity using the TCL master equation described in Sec.~\ref{sec:level2} for pulses implemented with DRAG on transmon qubits. 
We approximate the anharmonic oscillator as a three-level system (qutrit) and write the effective interaction frame Hamiltonian to second order in the Magnus expansion,
\begin{eqnarray}
\label{qutriteff}
    \tilde V_{\textrm{eff}}(t)=\tilde{\mathbf{r}}(t)\cdot\mathbf{\Lambda},
\end{eqnarray}
with $\mathbf \Lambda$ collecting the eight Gell-Mann matrices. The first order Magnus expansion amounts to four components in $\tilde{\mathbf r}$ corresponding to two symmetric and two anti-symmetric Gell-Mann matrices, while the second order Magnus term yields the components for the remaining symmetric and anti-symmetric matrices as well as for the diagonal ones, see the Appendix~\ref{app:qutrit}. Higher-order odd terms correct the first four components, while the even terms correct the other four components. We note that in the harmonic limit $(\Delta_a\to0)$, the second-order Magnus expansion yields the exact dynamics.

Integrating Eq.~\eqref{qutriteff} from $t-\tau$ to $t$ yields the vector $\mathbf r(t,\tau)$ for Eq.~\eqref{unitarymatrix1}. The eigenvalues of $\mathbf r\cdot\mathbf\Lambda$ are determined by the secular equation,
\begin{eqnarray}
    \det(\mathbf r\cdot\mathbf \Lambda-\mu)=-\mu^3+r^2\mu+\varepsilon=0,
\end{eqnarray}
with $r=|\mathbf r|$ and $\varepsilon=\det(\mathbf r\cdot\mathbf\Lambda)=\textrm{Tr}((\mathbf r\cdot\mathbf\Lambda)^3)/3$. The solution of the cubic secular equation is obtained using the trigonometric formulas by Vieta,
\begin{eqnarray}
\label{viete}
    \mu_j(\mathbf r)=\frac{2r}{\sqrt 3}\cos\left(\frac{1}{3}\arccos\left(\frac{3\sqrt{3}\varepsilon}{2r^3}\right)-\frac{2\pi j}{3}\right),
\end{eqnarray}
with $j=0,1,2$. These generate the characteristic function $K(\mathbf r)$ that is used to determine the renormalization of the time-dependent rates of the decay processes.
\begin{figure*}
    \centering
    \includegraphics[width=1\linewidth]{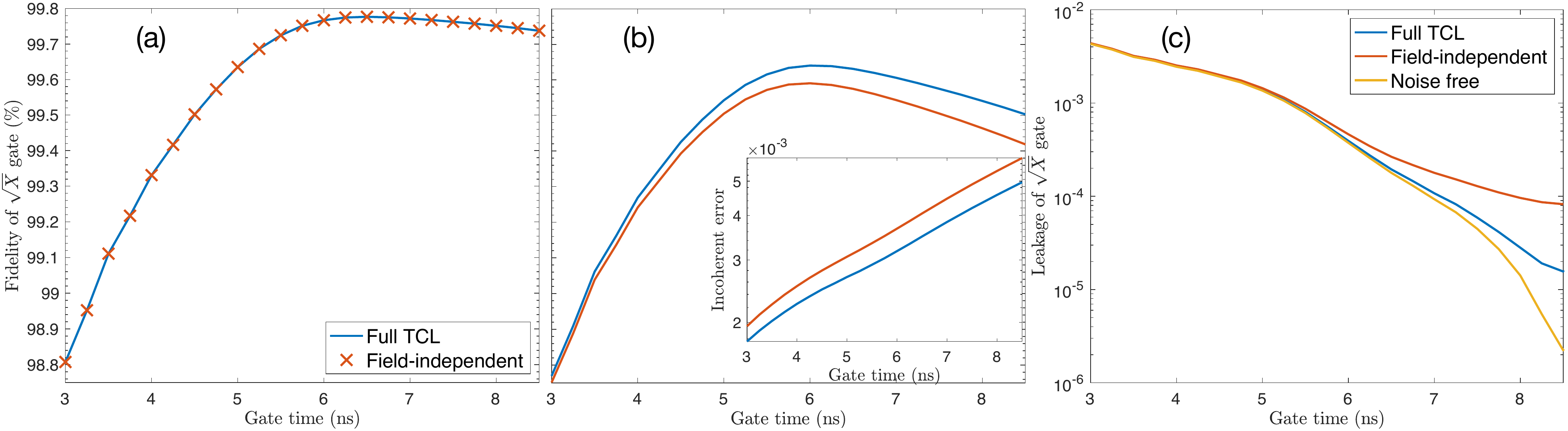}
    \caption{The average gate fidelity and average leakage outside the computational subspace as the function of gate time of an $X_{\pi/2}$ pulse or $\sqrt{X}$ gate, implemented with optimized DRAG technique on a transmon qubit. The blue curves correspond to the results obtained from the full TCL master equation, and the red curves to the results from the Redfield equation within the field-independent approximation. (a) In the presence of only  Ohmic relaxation, the field-independent picture captures the gate errors very well. (b) With the addition of dephasing, the field-dependent correlated effects become more relevant. For $1/f$ dephasing specifically these correlated effects increase the average gate fidelity compared to the field-independent model. The inset shows the infidelities without the coherent contribution, the full TCL solution containing every correlated effect shows lower incoherent gate error than the prediction of the Redfield dissipator within the field-independent approximation. (c) The increase in the average fidelity is accompanied by a decrease in the average leakage rate. The field-dependent correlated effects almost cancel the incoherent leakage. This is seen from the yellow curve, which shows the average leakage rate calculated using only the Hamiltonian contribution in the master equations (no environmental noise). The dimensionless noise strengths for these plots are $\lambda_R=5\cdot10^{-6}$ for Ohmic relaxation corresponding to $T_1\sim 3\ \mu s$ and $\lambda_\varphi=5\cdot10^{-8}$ corresponding to $T^{}_\varphi\sim 10\ \mu s$. The rest of the parameters: for the bosonic bath the inverse temperature is $\beta\omega_q=24$, high-frequency cutoff is $\omega_c=10\omega_q$; for dephasing the low-frequency cutoff is $\omega_{\textrm{ir}}=5\cdot10^{-5}\omega_q$ corresponding to $\sim600$ ns measurement time.}
    \label{Fidfig}
\end{figure*}

\subsection{Environmental effects and master equation}

Among the typical noise sources of Josephson-based qubits are the quasiparticles tunneling through the junction, leading to relaxation and dephasing \cite{catelani2011quasiparticle,catelani2012decoherence}, the electromagnetic environment causing charge and flux noise \cite{ithier2005decoherence}, spurious two-level systems that couple to the charge degree of freedom causing relaxation \cite{mull1,Muller2019}, and dispersively coupled two-level fluctuators causing dephasing \cite{transmonoise}. 

Here, we model the environmental effects independently and imagine that some entities with correlation function $C_Q(t)$ couple to the charge operator of the transmon $A_Q=a+a^\dagger$. A different bath with correlation function $C_\varphi(t)$ couples  to the excitation number operator $A_\varphi=a^\dagger a$, so that in Eq.~\eqref{filtering} describing the filtering operation, we have $C_{\alpha\beta}=\delta_{\alpha\beta}C_\alpha$. Due to the independence of the various noise sources, we are able to write the total dissipator as $\tilde D_Q+\tilde D_\varphi$ without any mixing between relaxation and dephasing.

We first discuss the dephasing dissipator. Since the interaction-frame operator is time-independent, $\tilde A_\varphi(t)=a^\dagger a$, its filtered version is simply $\tilde A_\varphi^{(f)}(t)=F_{\varphi,t}(a^\dagger a)$. Hence, the dephasing dissipator may be written as,
\begin{eqnarray}
\label{qutrit:deph}
    \tilde D_\varphi=\gamma_\varphi(t)\left(a^\dagger a\tilde\rho(t)a^\dagger a-\frac{1}{2}\{a^\dagger a a^\dagger a,\tilde\rho(t)\}\right)+\tilde D_\varphi^{(C)},\nonumber\\
\end{eqnarray}
with a renormalized dephasing rate 
\begin{eqnarray}
    \gamma_\varphi(t)=\frac{2}{N^2}\int_0^{t}\textrm d\tau\ \textrm{Re}(C_\varphi(\tau))|K(\mathbf r)|^2.
\end{eqnarray}
Within the qutrit approximation $N=3$, the characteristic function is obtained from Eq.~\eqref{viete} and the correction dissipator $\tilde D_\varphi^{(C)}$ is calculated from the general theory in Sec.~\ref{sec:level2} using $a^\dagger a=|1\rangle\langle 1|+2|2\rangle\langle 2|$. We assume that dephasing is caused by two-level fluctuators giving rise to $1/f$ noise. The correlation function is $C_\varphi(t)=\lambda_\varphi E_1(\omega_{\textrm{ir}}t)$, where $\omega_{\textrm {ir}}$ is a low-frequency cutoff determined by the measurement time, $\lambda_\varphi$ denotes the noise strength, and $E_1(x)$ is the exponential integral function \cite{guido1f}.

The charge operator in the interaction frame is $\tilde A_Q(t)=\tilde a(t)+\tilde a^\dagger(t)$ and within the qutrit approximation the creation operator is
\begin{eqnarray}
    \tilde a^\dagger(t)=e^{i\omega_qt}|1\rangle\langle0|+e^{i(\omega_q+\Delta_a)t}\sqrt{2}|2\rangle\langle1|.
\end{eqnarray}
We define the matrices $\Pi_1=|1\rangle\langle0|$ and $\Pi_2=|2\rangle\langle1|$ and express the dissipator as
\begin{gather}
    \tilde D_Q=\sum_{i,j=1}^2d^{(-)}_{ij}(t)\left(\Pi_i\tilde\rho(t)\Pi_j^\dagger-\frac{1}{2}\{\Pi_j^\dagger\Pi_i,\tilde\rho(t)\}\right)+\nonumber\\
    \sum_{i,j=1}^2d^{(+)}_{ij}(t)\left(\Pi^\dagger_i\tilde\rho(t)\Pi_j-\frac{1}{2}\{\Pi_j\Pi_i^\dagger,\tilde\rho(t)\}\right)+\tilde D_Q^{(C)},\label{qutrit:relax}
\end{gather}
where the decoherence matrices $d^{(\pm)}_{ij}(t)$ contain the renormalized time-dependent emission and absorption rates
and $\tilde D_Q^{(C)}$ contains the non-secular terms from the Redfield part, e.g., $\Pi_{1,2}\tilde\rho(t)\Pi_{1,2}$, as well as the correction terms from the drive, see Appendix~\ref{app:qutrit} for more details. For relaxation, we assume a bosonic bath with Ohmic spectral density. The dissipator in Eq.~\eqref{qutrit:relax} without the correction terms leads to the standard Lindblad dissipator for cascading relaxation within the Markovian approximation, which was used in previous studies \cite{Ex2,fastgate}. 

\subsection{Single qubit gate fidelity and leakage}

We wish to characterize how the correlated effects affect the relevant performance metrics of a single qubit gate operation on transmon qubits. 
First, we set the parameters of the transmon qubit to state-of-the-art values, with the qubit frequency $\omega_q/2\pi=5$ GHz, anharmonicity $\Delta_a/2\pi = -300$ MHz and numerically optimize the parameter $\xi$ appearing in the DRAG pulse shape. The optimization is done using the closed dynamics only, for fixed gate times ranging from 3 to 8.5 nanoseconds such that the gate fidelity is maximized. We choose the standard deviation in the Gaussian pulse shape as $\sigma=t_g/4$. These data are necessary to calculate the filtered jump operators.

We numerically solve the full TCL master equation in Eq.~\eqref{ME} using the dissipators in Eq.~\eqref{qutrit:deph} and Eq.~\eqref{qutrit:relax}. We use its solution to calculate the average single qubit gate fidelity and average leakage, defined in Appendix~\ref{sec:fid}, for a $\sqrt{X}$ gate. We refer to these results as the correlated solutions because these contain every temporally correlated effect within the weak coupling theory, including the effects of the field-dependent terms appearing in the dissipators of Eq.~\eqref{qutrit:deph} and Eq.~\eqref{qutrit:relax}.
We compare these results to the ones obtained using the Redfield dissipator within the field-independent approximation, which we refer to as the uncorrelated solution, as it ignores the correlations that arise due to the drive terms in the dissipators. 
Representative results are depicted in Fig.~\ref{Fidfig}. As expected from Eq.~\eqref{errorrelax}, including only the Ohmic relaxation in the environmental effects, the calculated average fidelity in the correlated and uncorrelated pictures agree very well, regardless of the relaxation noise strength. The additional correlations of $1/f$ dephasing noise manifest themselves as a visible shift in the average gate fidelity between the correlated and uncorrelated descriptions. Based on Eq.~\eqref{gatefid}, this shift should be positive so there should be an increase in the fidelity or a decrease in the gate error compared to the field-independent model because the noise power spectrum of $1/f$ noise decreases for increasing frequency. We find that this is indeed the case. Moreover, the shift is more pronounced for increasing gate times, this is because the unitary errors are dominant for gate times below $\sim10/\Delta_a$, which would correspond to roughly 5 ns for the chosen parameters. 

The increase in the average fidelity is accompanied by a decrease in the average rate of leakage outside the computational subspace. For increasing gate times, the difference between the correlated and field-independent descriptions can reach one order of magnitude. This suggests that the field-dependent, correlated effects may significantly contribute to the reduction of leakage in weakly anharmonic superconducting qubits. The trend of the metrics is evident in Fig.~\ref{Fidfig}. We report further numerical values using state-of-the-art parameters in Table \ref{Fidleak}.

\begin{table}[h]
\caption{\label{Fidleak}%
Performance metrics of an optimized $\sqrt{X}$ gate using DRAG, with 8 ns gate time. The dimensionless noise strengths are $\lambda_R=2.5\cdot10^{-7}$ corresponding to $T_1\sim 60\ \mathrm{\mu s}$ and $\lambda_\varphi=2.5\cdot10^{-9}$ corresponding to $T^{}_\varphi\sim 100\ \mathrm{\mu s}$.}
\begin{ruledtabular}
\begin{tabular}{rrr}
&\multicolumn{2}{c}{Error contribution to} \\
&Gate error&\multicolumn{1}{c}{Leakage}\\
\hline
Unitary\footnote{obtained by solving the closed dynamics}&$1.76\cdot10^{-5}$&$1.41\cdot10^{-5}$\\
Uncorrelated\footnote{obtained by solving the Redfield equation and by subtracting the unitary contribution}&$2.65\cdot10^{-4}$&$4.13\cdot10^{-6}$ \\ 
Correlated\footnote{remaining contribution to the full TCL result}
  &$-3.88\cdot10^{-5}$
  & $-3.41\cdot10^{-6}$ \\
  \hline
Total\footnote{contains all error sources, obtained from solving the full TCL}&$2.44\cdot10^{-4}$&$1.48\cdot10^{-5}$\\
\end{tabular}
\end{ruledtabular}
\end{table}





\section{\label{sec:level5}Conclusion}
Within the TCL framework of open quantum systems, we have established a quantum master equation capable of describing the effects of weakly coupled, temporally correlated quantum noise acting on arbitrarily driven quantum systems. By utilizing only the weak coupling assumption between the relevant system and the environment, the method discussed here can account for the renormalization of the well-known decay effects that affect the undriven system along with the additionally appearing corrections due to the presence of driving. 

The dissipation superoperator (dissipator) in the master equation which effectively characterizes the non-Hermitian influence of the environment consists of a system operator appearing in the system-environment interaction as well as a filtered version of the same operator. The main outcome of the presented method is the explicit evaluation of this filtered jump operator that leads to a description of every active decay channel and its time-dependent rate. We have illustrated and detailed our method on a generally driven two-level system and showed how the extra dissipation channels change the physics of a Rabi-driven qubit under the effect of decoherence compared to the field-independent approach, e.g., by producing a distinct non-equilibrium steady state.

Our model is well suited for the characterization of different noise processes that may degrade or even improve the performance of a quantum processor. In particular, we have shown that depending on the monotonicity of the noise power spectrum at low frequencies, the correlated effects may increase single qubit gate fidelities. We detailed this effect on a single qubit $\sqrt{X}$ gate implemented for a superconducting transmon qubit using DRAG pulses. Moreover, we also found that temporally correlated quantum noise may decrease the average leakage rate, the probability of undesired transitions from the computational subspace to other levels of the quantum system. Admittedly, these results are subtle but quite relevant in the context of quantum error correction, which requires the physical error rate per quantum operation to be below a threshold that enables the correction of errors faster than they occur. The gate failure threshold depends on the type of the code, alongside various parameters including the memory failure rate, the physical scale-up of the computer size, and the time required for measurements and classical processing \cite{steane}. As a result, the error threshold may vary between $10^{-3}$ and $10^{-6}$. At these levels, the presented correlated effects are indeed significant.

\begin{acknowledgments}
We are grateful to Joris Kattem\"olle for insightful discussions. We wish to acknowledge the support from the German Ministry for Education and Research, under the QSolid project, Grant No.~13N16167.
\end{acknowledgments}

\appendix

\section{\label{magna}The Magnus expansion and the effective Hamiltonian}

The purpose of this section is to summarize the properties of the Magnus expansion \cite{magnus1} relevant to Sec.~\ref{subs:mag} and derive the effective drive Hamiltonian appearing in Eq.~\eqref{formalexact}. In the filtering operation the closed dynamics is represented by $\tilde U(t,t-\tau)=\tilde U(t,t_0)\tilde U^\dagger(t-\tau,t_0)$, which can be constructed by the solution of Eq.~\eqref{UI}. The solution is written as $ \tilde U(t,t_0)=e^{M(t,t_0)}$, with the Magnus operator given by the series,
\begin{equation}
    M(t,t_0) = \sum_{n=1}^\infty M_{n}(t,t_0).
\end{equation}
The terms in the Magnus series are obtained from the recursive formula,
\begin{gather}
    M_1(t,t_0)=-i\int_{t_0}^t\textrm dt_1\ \tilde V(t_1),\nonumber\\
    S_k^{(1)}(t)=-i[M_{k-1}(t),\tilde V(t)],\ k\geq2,\nonumber\\
    S_k^{(j)}(t)=\sum_{m=1}^{k-j}[M_{m}(t),S_{k-m}^{(j-1)}(t)],\nonumber\\
     M_n(t,t_0)=\sum_{k=1}^{n-1}\frac{B_k}{k!}\int_{t_0}^t\textrm dt_1\ S_n^{(k)}(t_1),\ n\geq2,\label{magrec}
\end{gather}
where $B_k$ denotes the Bernoulli numbers. In general, the Magnus expansion only converges if $\tilde V(t)$ is ``small". More precisely, the series is absolute convergent for $t_0<t\leq T$, with
\begin{equation}
    T=\textrm{max}\left(t\geq t_0:\ \int_{t_0}^t\textrm dt_1\ ||\tilde V(t_1)||_2<R\right),
\end{equation}
where $||\tilde V||_2$ is the spectral norm characterized by the root of the largest eigenvalue of $\tilde V^\dagger\tilde V$  and the radius of convergence is $R\approx1.0868$ \cite{Blanes1998}. To second order the Magnus operator is
\begin{equation}
iM(t,t_0)=\int_{t_0}^t\mathrm dt_1\ \left(\tilde V(t_1)-\frac{i}{2}\int_{t_0}^{t_1}\mathrm dt_2\ \left[\tilde V(t_2),\tilde V(t_1)\right]\right),    
\end{equation}
from which we can read off the second order effective drive Hamiltonian $\tilde V_{\textrm{eff}}(t,t_0)$. As noted, $\tilde V_{\textrm{eff}}$ explicitly depends on the initial time $t_0$; to find $\tilde U(t,t-\tau)$ the initial time is replaced by $t-\tau$ in Eqs.~\eqref{magrec} and we obtain the expression in Eq.~\eqref{veff}. 

We remark that if the interaction-frame drive Hamiltonian has the form $\tilde V(t)=f(t)\hat O$, where $f(t)$ is a continuous function of time and $\hat O$ is a time-independent system operator, then the effective Hamiltonian is simply $V_{\textrm{eff}}(t,\tau)=\tilde V(t)$. Such a form appears, e.g., for a resonantly driven two-level system within the RWA, as in Sec.~\ref{rabi}.

\section{\label{hugeformulas}Details for the correction dissipators}

Using the recurrence relation of the Magnus expansion and the properties of the generalized Gell-Mann matrices, we were able to expand the unitary $\tilde U(t,t-\tau)$. Substituting the expression of the unitary \eqref{unitarymatrix2} into Eq.~\eqref{filtering} leads to Eq.~\eqref{filtering2}, which we repeat here,
\begin{equation*}
F_{\alpha\beta,t}(\hat P_\textbf{n})=\tilde\Gamma_{\alpha\beta}(\omega_{\textbf n}, t)\hat P_\textbf{n}+\hat M_{\alpha\beta}(\omega_\textbf n,t).
\end{equation*}
The first term corresponds to the renormalization of the rates appearing in Eq.~\eqref{eq:19}, and it is obtained using the diagonal matrices in the expression of the unitary in Eq.~\eqref{unitarymatrix2}. The second term is the correction matrix that does not have any matrix element proportional to $\hat P_\textbf{n}$. This matrix depends only on the derivatives of the characteristic function, and it is given by
\begin{gather}
\hat M_{\alpha\beta}(\omega_{\textbf n},t)=\sum_{k=1}^{N(N-1)}\lambda_{\alpha\beta,k}^{(1)}(\omega_{\textbf n},t)\hat P_\textbf{n}\Lambda_k\nonumber\\
+\sum_{k=1}^{N(N-1)}\lambda_{\alpha\beta,k}^{(2)}(\omega_{\textbf n},t)\Lambda_k\hat P_\textbf{n}\nonumber\\
+\sum_{k=1}^{N(N-1)}\sum_{l=1}^{N^2-1}\lambda_{\alpha\beta,kl}^{(3)}(\omega_{\textbf n},t)\Lambda_k\hat P_\textbf{n}\Lambda_l\nonumber\\
+\sum_{k=N(N-1)+1}^{N^2-1}\sum_{l=1}^{N(N-1)}\lambda_{\alpha\beta,kl}^{(3)}(\omega_{\textbf n},t)\Lambda_k\hat P_\textbf{n}\Lambda_l,\label{correction}
\end{gather}
with generalized (complex) rates
\begin{align}
\lambda_{\alpha\beta,k}^{(1)}=-\frac{i}{2N}\int_{0}^{\Delta t}\mathrm d\tau\ C_{\alpha\beta}(\tau)e^{-i\omega_{\textbf n}\tau}K(\mathbf r)\partial_kK^*(\mathbf r),   \nonumber\\
\lambda_{\alpha\beta,k}^{(2)}=\frac{i}{2N}\int_{0}^{\Delta t}\mathrm d\tau\ C_{\alpha\beta}(\tau)e^{-i\omega_{\textbf n}\tau}K^*(\mathbf r)\partial_kK(\mathbf r),   \nonumber\\
\lambda_{\alpha\beta,kl}^{(3)}=\frac{1}{4}\int_{0}^{\Delta t}\mathrm d\tau\ C_{\alpha\beta}(\tau)e^{-i\omega_{\textbf n}\tau}\partial_kK(\mathbf r)\partial_lK^*(\mathbf r).  \nonumber\\
\label{genrate}
\end{align}
In the first two rows of Eq.~\eqref{correction}, the summation runs through only the symmetric and anti-symmetric components of the vector $\mathbf{\Lambda}$, respectively. In the third row the diagonal components are absent from the sum running over the first index, in the fourth row the sum over the first index runs through only the diagonal components and the second index runs through only the symmetric and anti-symmetric components. 

As mentioned above, the renormalized rates are obtained using only the diagonal part of the unitary; as such they have two parts: $\tilde\Gamma_{\alpha\beta}(\omega_{\textbf n}, t)=\tilde\Gamma_{\alpha\beta}^{(1)}(\omega_{\textbf n}, t)+\tilde\Gamma_{\alpha\beta}^{(2)}(\omega_{\textbf n}, t)$. The first term is calculated using the unit matrix and it depends on the absolute square of the characteristic function,
\begin{eqnarray}
\label{gammaab}
\tilde\Gamma_{\alpha\beta}^{(1)}(\omega_{\textbf n}, t)=\int_{0}^{\Delta t}\mathrm d\tau\ C_{\alpha\beta}(\tau)e^{-i\omega_{\textbf n}\tau}\frac{|K(\mathbf r)|^2}{N^2}.
\end{eqnarray}
To arrive at the other term, we multiply $\hat P_\textbf{n}$ from the left or right with the diagonal Gell-Mann matrices, which yields
\begin{gather}
\hat P_\textbf{n}\Lambda_l^d=\left(\sqrt{\frac{2}{l(l+1)}}\sum_{j=0}^{l-1}\delta_{mj}-\sqrt{\frac{2l}{(l+1)}}\delta_{ml}\right)\hat P_\textbf{n},\nonumber\\
\Lambda_l^d\hat P_\textbf{n}=\left(\sqrt{\frac{2}{l(l+1)}}\sum_{j=0}^{l-1}\delta_{nj}-\sqrt{\frac{2l}{(l+1)}}\delta_{nl}\right)\hat P_\textbf{n}.\nonumber
\end{gather}
These expressions are used to calculate precisely the terms that are absent from the sums
of the correction matrix $\hat M_{\alpha\beta}$ as these are proportional to $\hat P_\textbf{n}$. The remaining parts of the rate renormalization are written as
\begin{gather}
\tilde\Gamma_{\alpha\beta}^{(2)}(\omega_{\textbf n}, t)=\sum_{l=1}^{N-1}c_{ml}\lambda_{\alpha\beta,N(N-1)+l}^{(1)}+c_{nl}\lambda_{\alpha\beta,N(N-1)+l}^{(2)}\nonumber\\
+\sum_{k,l=1}^{N-1}c_{nk}c_{ml}\lambda_{\alpha\beta,N(N-1)+k,N(N-1)+l}^{(3)}.\label{deltaab}
\end{gather}
Here, $c_{ml}$ corresponds to the coefficient appearing in the formula for $\hat P_\textbf{n}\Lambda_l^d$, and $c_{nl}$ corresponds to the coefficient appearing in the formula for $\Lambda_l^d\hat P_\textbf{n}$.

\section{\label{relaxapp}Relaxation dissipator for the generally driven qubit}

In this section of the Appendix, we detail the derivation of the decoherence matrix appearing in Eq.~\eqref{Drelax}. The transverse coupling to the environment $(A_R=\sigma_x)$ leads to the interaction-frame operator $\tilde\sigma_x(t)=e^{i\omega_qt}\sigma_-+e^{-i\omega_qt}\sigma_+$. The filtered operator entering the master equation is $\tilde\sigma_x^{(f)}(t)=e^{i\omega_qt}F_t(\sigma_-)+e^{-i\omega_qt}F_t(\sigma_+)$, where we need to calculate the filtering operations for $\sigma_\mp$. This is a straightforward substitution into the expressions of Appendix~\ref{hugeformulas}, using the characteristic function $K(\mathbf r)=2\cos r$. As an example, we detail the derivation for $F_t(\sigma_-)$.

According to Eq.~\eqref{filtering2}, the filtering yields $F_t(\sigma_-)=\tilde\Gamma(\omega_q,t)\sigma_-+\hat M(\omega_q,t)$. The two parts of the renormalized rates are
\begin{gather}
    \tilde\Gamma^{(1)}(\omega_q,t)=\int_{0}^{\Delta t}\mathrm d\tau\ C_{}(\tau)e^{-i\omega_{q}\tau}\cos^2r,\nonumber\\
    \tilde\Gamma^{(2)}(\omega_q,t)=c_{01}\lambda_{\ 3}^{(1)}+c_{11}\lambda_{\ 3}^{(2)}+c_{11}c_{01}\lambda_{\ 33}^{(3)}=\nonumber\\
    \int_{0}^{\Delta t}\mathrm d\tau\ C_{}(\tau)e^{-i\omega_{q}\tau}\left(\frac{ir_z}{r}\sin2r-\frac{r_z^2}{r^2}\sin^2r\right),
\end{gather}
where we used $\sigma_-=|n\rangle\langle m|$ with $n=1$ and $m=0$, so that $c_{01}=1,c_{11}=-1$, and $\partial_3K(r)=\partial_{r_z}K(r)=-2r_z/r\sin r$. For qubits, the generalized Gell-Mann matrices are the Pauli matrices, these are ordered into a vector as $\mathbf\Lambda=(\sigma_x\ \sigma_y\ \sigma_z)^T$. The correction matrix is then obtained from Eq.~\eqref{correction}, which leads to
\begin{align}
    \hat M(\omega_q,t)=&\lambda_{\ 1}^{(1)}\sigma_-\sigma_x+\lambda_{\ 2}^{(1)}\sigma_-\sigma_y+\lambda_{\ 1}^{(2)}\sigma_x\sigma_-
    +  \lambda_{\ 2}^{(2)}\sigma_y\sigma_-\nonumber\\ 
   & +\lambda_{\ 11}^{(3)}\sigma_x\sigma_-\sigma_x+\lambda_{\ 12}^{(3)}\sigma_x\sigma_-\sigma_y++\lambda_{\ 13}^{(3)}\sigma_x\sigma_-\sigma_z\nonumber\\
    &+\lambda_{\ 21}^{(3)}\sigma_y\sigma_-\sigma_x+\lambda_{\ 22}^{(3)}\sigma_y\sigma_-\sigma_y+\lambda_{\ 23}^{(3)}\sigma_y\sigma_-\sigma_z\nonumber\\
    &+\lambda_{\ 31}^{(3)}\sigma_z\sigma_-\sigma_x+\lambda_{\ 32}^{(3)}\sigma_z\sigma_-\sigma_y.\label{corrM}
\end{align}
For the driven qubit, the characteristic function is real, and hence the general rates obey the relations $\lambda_{\ k}^{(1)}=-\lambda_{\ k}^{(2)}$ and $\lambda_{\ kl}^{(3)}=\lambda_{\ lk}^{(3)}$. After the evaluation of Eq.~\eqref{corrM} we find
\begin{align}
    \hat M(\omega_q,t)= \int_{0}^{\Delta t}\mathrm d\tau\ C_{}(\tau)e^{-i\omega_{q}\tau}\left(\frac{r_x-ir_y}{r}\sin r\right)^2\sigma_+\nonumber\\
    +\int_{0}^{\Delta t}\mathrm d\tau\ C_{}(\tau)e^{-i\omega_{q}\tau}\frac{r_x-ir_y}{r}\left(\frac{r_z}{r}\sin^2r-\frac{i}{2}\sin2r\right)\sigma_z.
\end{align}
A similar calculation leads to the filtering of $\sigma_+$, with the renormalized rate
\begin{gather}
    \tilde\Gamma^{(1)}(-\omega_q,t)=\int_{0}^{\Delta t}\mathrm d\tau\ C_{}(\tau)e^{i\omega_{q}\tau}\cos^2r,\nonumber\\
    \tilde\Gamma^{(2)}(-\omega_q,t)=
    -\int_{0}^{\Delta t}\mathrm d\tau\ C_{}(\tau)e^{i\omega_{q}\tau}\left(\frac{ir_z}{r}\sin2r+\frac{r_z^2}{r^2}\sin^2r\right),
\end{gather}
and correction matrix
\begin{align}
    \hat M(-\omega_q,t)= \int_{0}^{\Delta t}\mathrm d\tau\ C_{}(\tau)e^{i\omega_{q}\tau}\left(\frac{r_x+ir_y}{r}\sin r\right)^2\sigma_-\nonumber\\
    +\int_{0}^{\Delta t}\mathrm d\tau\ C_{}(\tau)e^{i\omega_{q}\tau}\frac{r_x+ir_y}{r}\left(\frac{r_z}{r}\sin^2r+\frac{i}{2}\sin2r\right)\sigma_z.
\end{align}
Substituting all of the above into Eq.~\eqref{dissipator1} yields the dissipator for a generally driven qubit under relaxation. The terms are rearranged as in Eq.~\eqref{Drelax} with the decoherence matrix elements $\gamma_{ij}(t)$ reported below. The renormalized emission and absorption rates are the first two diagonal elements, with the relabeling $\gamma_{\pm\pm}\equiv\gamma_\pm$,
\begin{align}\label{emrate}
    \gamma_\pm(t)= &\tilde\Gamma^{}(\mp\omega_q,t)+ \tilde\Gamma^{*}(\mp\omega_q,t)\nonumber\\
   &+ e^{\pm 2 i\omega_qt}\textrm{Tr}\left[\sigma_\mp\hat M(\pm\omega_q,t)\right]\nonumber\\
   &+e^{\mp 2 i\omega_qt}\textrm{Tr}\left[\sigma_\pm\hat M^\dagger(\pm\omega_q,t)\right].
\end{align}
The non-secular rates are the first elements in the first superdiagonal (subdiagonal), with $\gamma_{+-}\equiv\gamma_{\textrm{ns}}$ ($\gamma_{-+}=\gamma_{+-}^*$), 
\begin{gather}
    \gamma_{\textrm{ns}}(t)=\textrm{Tr}\left[\sigma_-(\hat M(\omega_q,t)+\hat M^\dagger(-\omega_q,t))\right]\nonumber\\
    + e^{-2 i\omega_qt}\left( \tilde\Gamma^{}(-\omega_q,t)+ \tilde\Gamma^{*}(\omega_q,t)\right).
\end{gather}
Finally, the rates originating purely from the correction terms due to the drive are $\gamma_{z+}=\gamma_{+z}^*$ and $\gamma_{z-}=\gamma_{-z}^*$; these read
\begin{gather}
    \gamma_{z\pm}(t)=\textrm{Tr}\left[\sigma_z(\hat M(\mp\omega_q,t)+e^{\pm2 i\omega_qt}\hat M(\pm\omega_q,t))\right].
\end{gather}
Some of these expressions are not independent; we note that the following identities hold
\begin{align}
    \gamma_+(t)+\gamma_-(t) &= e^{2 i\omega_qt}\gamma_{\textrm{ns}}(t)+e^{-2 i\omega_qt}\gamma^*_{\textrm{ns}}(t),\nonumber\\
    \gamma_{z+}(t)&= e^{2 i\omega_qt}\gamma_{z-}(t).\label{identities}
\end{align}
From these identities, it follows that the decoherence matrix $\gamma_{ij}(t)$ in Eq.~\eqref{Drelax} has only two non-zero eigenvalues,
\begin{align}
    d_{1,2}(t)=\frac{\gamma_+(t)+\gamma_-(t)}{2}\nonumber\\
    \pm\sqrt{\frac{(\gamma_+(t)-\gamma_-(t))^2}{4}+4|\gamma_{z-}(t)|^2+|\gamma_{ns}(t)|^2},
\end{align}
where $d_2(t)$ may take negative values corresponding to non-Markovian evolution, and $d_3(t)=0$. As a result, the dissipator for arbitrary driving may be rewritten into the canonical Lindblad form with only two decay channels corresponding to the canonical rates $d_{1,2}(t)$ and jump operators $L_{1,2}(t)$.

\section{\label{sec:fid}Performance metrics for the characterization of gate operations: average gate fidelity and average leakage}

For a particular quantum operation implemented with a control protocol $V(t)$, we define the actual realization of the operation through the solution of Eq.~\eqref{ME} which is subsequently projected onto the computational subspace. The realization corresponds to the mapping,
\begin{eqnarray}
\mathcal E=\mathcal P\tilde\rho(t_f)\mathcal P,
\end{eqnarray}
where $t_f$ is the final time instant of the operation and $\mathcal P$ is the operator projecting onto the computational subspace. In general, due to the projection, the realization $\mathcal E$ is a trace-non-preserving quantum map.

To estimate the performance of a protocol that tries to implement an ideal quantum operation $\mathcal U$, we refer to the average fidelity between the realization and the target operation, given by the formula valid for trace-non-preserving maps \cite{fidnonTP},
\begin{eqnarray}
\label{fidform}
F(\mathcal E,\mathcal U)=\frac{d}{d+1}F_e+\frac{1}{d+1}\textrm{Tr}\left(\mathcal E\left(\frac{\mathbb 1}{d}\right)\right).
\end{eqnarray}
Here, $d$ is the dimension of the computational subspace, $\mathbb 1$ corresponds to the identity in the computational subspace and $F_e=\langle\phi|\mathbb 1\otimes(\mathcal U^\dagger\circ\mathcal E)(\phi)|\phi\rangle$ is the entanglement fidelity with $\phi$ being a maximally entangled state in the doubled system Hilbert space \cite{fid0,fid1,fid2}. The second term in Eq.~\eqref{fidform} represents the effects of leakage on the performance of the realization $\mathcal E$, as we identify the action of $\mathcal E$ on the maximally mixed state with the average leakage, defined in Ref.~\cite{Gambetta2018}, 
\begin{eqnarray}
\label{leak}
L=1-\textrm{Tr}\left(\mathcal E\left(\frac{\mathbb 1}{d}\right)\right).
\end{eqnarray}

\begin{widetext}

\section{\label{app:qutrit}Transmon qutrit effective Hamiltonian and master equation}
The components of the vector appearing in the effective drive Hamiltonian  $\tilde V_{\textrm{eff}}$ in Eq.~\eqref{qutriteff} are as follows,
\begin{eqnarray}
    \tilde r_1 &=& \Omega(t)\cos\omega_qt,\quad \tilde r_2 = \sqrt{2}\Omega(t)\cos(\omega_q+\Delta_a)t,\quad \tilde r_4 = \Omega(t)\sin\omega_qt, \quad \tilde r_5 = \sqrt{2}\Omega(t)\sin(\omega_q+\Delta_a)t,\nonumber\\
    \tilde r_3 &=& \sqrt{2}\Omega(t)\int_{t-\tau}^t\textrm dt_1\ \Omega(t_1)\sin\left(\frac{\Delta_a}{2}(t-t_1)\right)\cos\left(\left(\omega_q+\frac{\Delta_a}{2}\right)(t+t_1)\right),\nonumber\\
    \tilde r_6 &=& \sqrt{2}\Omega(t)\int_{t-\tau}^t\textrm dt_1\ \Omega(t_1)\sin\left(\frac{\Delta_a}{2}(t-t_1)\right)\sin\left(\left(\omega_q+\frac{\Delta_a}{2}\right)(t+t_1)\right),\nonumber\\
    \tilde r_7 &=& 2\Omega(t)\int_{t-\tau}^t\textrm dt_1\ \Omega(t_1)\sin\left(\frac{\Delta_a}{2}(t-t_1)\right)\cos\left(\left(\omega_q+\frac{\Delta_a}{2}\right)(t-t_1)\right),\nonumber\\
    \tilde r_8 &=& -\sqrt{3}\Omega(t)\int_{t-\tau}^t\textrm dt_1\ \Omega(t_1)\sin\left(\omega_q+\Delta_a\right)\left(t-t_1\right).
\end{eqnarray}
Integrating these over time yields the vector appearing in Sec.~\ref{sec:level2}, $\mathbf r(t,\tau)=\int_{t-\tau}^t\textrm dt_1\ \tilde r(t_1,\tau)$. Using Eq.~\eqref{viete}, the characteristic function $K(\mathbf r)$ is obtained.

The decoherence matrices appearing in the dissipator Eq.~\eqref{qutrit:relax} are
\begin{eqnarray}
    d^{(\mp)}_{ij}(t)=\begin{pmatrix}
    2\textrm{Re}\left(\tilde\gamma(\pm\omega_q,t)\right)&\sqrt{2}e^{\mp i\Delta_at}\left(\tilde\gamma(\pm\omega_q,t)+\tilde\gamma^*(\pm\omega_q\pm\Delta_a,t)\right)\\
    \sqrt{2}e^{\pm i\Delta_at}\left(\tilde\gamma^*(\pm\omega_q,t)+\tilde\gamma(\pm\omega_q\pm\Delta_a,t)\right)&4\textrm{Re}\left(\tilde\gamma(\pm\omega_q\pm\Delta_a,t)\right)
    \end{pmatrix},
\end{eqnarray}
with time-dependent rates obtained from Eq.~\eqref{gammaab} and Eq.~\eqref{deltaab},
\begin{eqnarray}
    \tilde\gamma(\omega,t)=\int_0^t\textrm d\tau\ C_Q(\tau)e^{- i\omega\tau}\left(\frac{|K(\mathbf r)|^2}{N^2}+f(\omega,\partial_kK(\mathbf r))\right),
\end{eqnarray}
where $f$ denotes the expression containing the derivatives of the characteristic function obtained from Eq.~\eqref{genrate} and Eq.~\eqref{deltaab}. For example, if $\omega=\omega_q$, the expression $f(\omega,\partial_kK(\mathbf r))$ is
\begin{eqnarray}
    f(\omega_q,\partial_kK(\mathbf r))=\frac{i}{2N}
    \bigg(\frac{K^*(\mathbf r)\partial_8 K(\mathbf r)-K(\mathbf r)\partial_8 K^*(\mathbf r)}{\sqrt{3}}-(K^*(\mathbf r)\partial_7 K(\mathbf r)+K(\mathbf r)\partial_7 K^*(\mathbf r))\bigg)\nonumber\\
    +\frac{1}{4}\bigg(-\partial_7K(\mathbf r)\partial_7K^*(\mathbf r)-\frac{\partial_7K(\mathbf r)\partial_8K^*(\mathbf r)-\partial_8K(\mathbf r)\partial_7K^*(\mathbf r)}{\sqrt{3}}+\frac{1}{3} \partial_8K(\mathbf r)\partial_8K^*(\mathbf r)\bigg).
\end{eqnarray}
Similar expressions for $\omega=-\omega_q,\pm \omega_q\pm\Delta_a$ exist, but have been omitted.
\end{widetext}

\bibliography{leakage}

\end{document}